\documentclass[preprintnumbers,article,amsmath,amssymb,floatfix,10pt,prd,twocolumn,
superscriptaddress,nofootinbib]{revtex4-2}
\usepackage{bm}
\usepackage{amsfonts}
\usepackage{latexsym}
\usepackage[latin1]{inputenc}
\usepackage{graphicx}
\usepackage{amsmath}
\usepackage{bigints} 
\usepackage{palatino}
\usepackage{mathrsfs} 
\usepackage{mathpazo}
\usepackage{textcomp}
\usepackage{float}
\usepackage{booktabs}
\usepackage{dcolumn}
\usepackage{ragged2e}
\usepackage{hyperref}
\usepackage{relsize} 
\hypersetup{colorlinks,citecolor=blue}
\hypersetup{colorlinks=true,linkcolor=red,filecolor=magenta,    urlcolor=blue}
\usepackage{amsmath}
\usepackage{xcolor}
\usepackage{orcidlink}
\usepackage{epsfig}
\usepackage{subfigure}
\usepackage{commath}

\usepackage{titlesec}
\titleclass{\subsubsubsection}{straight}[\subsection]
\newcounter{subsubsubsection}[subsubsection]
\renewcommand\thesubsubsubsection{\thesubsubsection.\arabic{subsubsubsection}}
\titleformat{\subsubsubsection}
  {\normalfont\normalsize\bfseries}{\thesubsubsubsection}{1em}{}
\titlespacing*{\subsubsubsection}
{0pt}{3.25ex plus 1ex minus .2ex}{1.5ex plus .2ex}
\makeatletter
\renewcommand\paragraph{\@startsection{paragraph}{5}{\z@}%
  {3.25ex \@plus1ex \@minus.2ex}%
  {-1em}%
  {\normalfont\normalsize\bfseries}}
\renewcommand\subparagraph{\@startsection{subparagraph}{6}{\parindent}%
  {3.25ex \@plus1ex \@minus .2ex}%
  {-1em}%
  {\normalfont\normalsize\bfseries}}
\def\toclevel@subsubsubsection{4}
\def\toclevel@paragraph{5}
\def\toclevel@paragraph{6}
\def\l@subsubsubsection{\@dottedtocline{4}{7em}{4em}}
\def\l@paragraph{\@dottedtocline{5}{10em}{5em}}
\def\l@subparagraph{\@dottedtocline{6}{14em}{6em}}
\makeatother
\DeclareFontFamily{U}{BOONDOX-calo}{\skewchar\font=45 }
\DeclareFontShape{U}{BOONDOX-calo}{m}{n}{
  <-> s*[1.05] BOONDOX-r-calo}{}
\DeclareFontShape{U}{BOONDOX-calo}{b}{n}{
  <-> s*[1.05] BOONDOX-b-calo}{}
\DeclareMathAlphabet{\mathcalboondox}{U}{BOONDOX-calo}{m}{n}
\SetMathAlphabet{\mathcalboondox}{bold}{U}{BOONDOX-calo}{b}{n}
\DeclareMathAlphabet{\mathbcalboondox}{U}{BOONDOX-calo}{b}{n}

\def\jnl@style{\it}
\def\aaref@jnl#1{{\jnl@style#1}}

\def\aaref@jnl#1{{\jnl@style#1}}

\def\aj{\aaref@jnl{AJ}}                   
\def\apj{\aaref@jnl{ApJ}}                 
\def\apjl{\aaref@jnl{ApJ}}                
\def\apjs{\aaref@jnl{ApJS}}               
\def\apss{\aaref@jnl{Ap\&SS}}             
\def\aap{\aaref@jnl{A\&A}}                
\def\aapr{\aaref@jnl{A\&A~Rev.}}          
\def\aaps{\aaref@jnl{A\&AS}}              
\def\mnras{\aaref@jnl{Mon.~Not.~Roy.~Astron.~Soc.}}             
\def\prd{\aaref@jnl{Phys.~Rev.~D}}        
\def\prc{\aaref@jnl{Phys.~Rev.~C}}  
\def\prl{\aaref@jnl{Phys.~Rev.~Lett.}}    
\def\qjras{\aaref@jnl{QJRAS}}             
\def\skytel{\aaref@jnl{S\&T}}             
\def\ssr{\aaref@jnl{Space~Sci.~Rev.}}     
\def\zap{\aaref@jnl{ZAp}}                 
\def\nat{\aaref@jnl{Nature}}              
\def\aplett{\aaref@jnl{Astrophys.~Lett.}} 
\def\apspr{\aaref@jnl{Astrophys.~Space~Phys.~Res.}} 
\def\physrep{\aaref@jnl{Phys.~Rep.}}      
\def\physscr{\aaref@jnl{Phys.~Scr}}       
\def\commat{\aaref@jnl{Comm.~Math.~Phys.}}              
\def\science{\aaref@jnl{Science}}               
\def\cqg{\aaref@jnl{Classical Quant.~Grav.}}            
\def\jpcs{\aaref@jnl{JPCS}}                                     
\def\ijmpd{\aaref@jnl{Int.~J.~Mod.~Phys.~D}}                    
\def\grg{\aaref@jnl{Gen.~Relat.~Gravit.}}               
\def\rpp{\aaref@jnl{Rep.~Prog.~Phys.}}          
\def\npa{\aaref@jnl{Nucl.~Phys.~A}}        
\def\lrr{\aaref@jnl{Living Rev.~Rel.}}                   
\def\jcap{\aaref@jnl{J.~Cosmology Astropart.~Phys.}}    
\def\rmp{\aaref@jnl{Rev.~Mod.~Phys.}}   
\def\epjc{\aaref@jnl{Eur.~Phys.~J.~C}} 
\def\plb{\aaref@jnl{~Phy.~Lett.~B}} 
\def\mpla{\aaref@jnl{Mod.~Phy.~Lett.~A}} 
\def\arxiv{\aaref@jnl{arxiv.org}}

\allowdisplaybreaks[1]

\begin{document}
\color{black} 
\title{Exploring wormhole solutions with global monopole charge in the context of $f(Q)$ gravity}

\author{Moreshwar Tayde\orcidlink{0000-0002-3110-3411}}
\email{moreshwartayde@gmail.com}
\affiliation{Department of Mathematics, Birla Institute of Technology and
Science-Pilani,\\ Hyderabad Campus, Hyderabad-500078, India.}

\author{P.K. Sahoo\orcidlink{0000-0003-2130-8832}}
\email{pksahoo@hyderabad.bits-pilani.ac.in}
\affiliation{Department of Mathematics, Birla Institute of Technology and
Science-Pilani,\\ Hyderabad Campus, Hyderabad-500078, India.}

%
\date{\today}

\begin{abstract}
This study explores the potential existence of traversable wormholes influenced by a global monopole charge within the $f(Q)$ gravity framework. To elucidate the characteristics of these wormholes, we conducted a comprehensive analysis of wormhole solutions employing three different forms of redshift function under a linear $f(Q)$ model. Wormhole shape functions were derived for barotropic, anisotropic, and isotropic Equations of State (EoS) cases. However, in the isotropic EoS case, the calculated shape function failed to satisfy the asymptotic flatness condition. Additionally, we observed that our obtained shape functions adhered to the flaring-out conditions under an asymptotic background for the remaining EoS cases.
Furthermore, we examined the energy conditions at the wormhole throat with a radius $r_0$. We noted the influences of the global monopole's parameter $\eta$, the EoS parameter $\omega$, and $n$ in violating energy conditions, particularly the null energy conditions. Finally, we conducted a stability analysis utilizing the Tolman-Oppenheimer-Volkov (TOV) equation and found that our obtained wormhole solution is stable.
\end{abstract}

\maketitle



\section{Introduction}\label{sec1}
The wormhole concept was first proposed by L. Flamm \cite{L. Flamm} in 1916. Subsequently, Einstein and Rosen further explored the notion of a wormhole, introducing a hypothetical structure known as the Einstein-Rosen bridge \cite{A. Einstein}. Wormholes are theoretical constructs with topological properties that provide a conduit for connecting distinct regions of Space-time. These structures are envisioned as tube-like formations, asymptotically flat at both ends and connected by a throat. Wormholes are categorized based on the nature of their throats, distinguishing between static and non-static variants. A wormhole with a constant throat radius is termed a static wormhole, while a non-static wormhole exhibits a variable throat radius. Fuller and Wheeler demonstrated that traversing through the Einstein-Rosen bridge, even for a photon, would lead to its instantaneous collapse upon formation \cite{R. W. Fuller}. Subsequently, Morris et al. \cite{M. S. Morris 1} proposed that exotic forms of matter threaded through a wormhole might stabilize it; however, the feasibility of such requirements remains uncertain. Finally, Morris and Thorne \cite{M. S. Morris} presented static traversable wormholes, which offer a theoretical framework for interstellar travel and constitute exact solutions within the framework of General Relativity (GR). The authors demonstrated that a wormhole could be traversable if it possesses exotic matter with a minimal surface area that satisfies the flaring-out condition. This exotic matter is essential for constructing a wormhole in GR; indeed, in classical GR, wormhole solutions violate all energy conditions, especially null energy conditions \cite{M. Visser 1}. However, this type of hypothetical matter exhibits unusual properties within the framework of GR. In contrast, from the perspective of quantum gravity, it can arise as a natural consequence of fluctuations in the Space-time topology \cite{J. A. Wheeler}. Thus, minimizing the violation of the energy conditions or reducing the quantity of exotic matter at the throat is essential.\\
\indent Kanti et al. made important advances in the research of traversable wormholes by proposing wormhole constructs within the framework of quadratic gravitational theories \cite{P. Kanti 1, P. Kanti 2}. This unique technique provides an intriguing concept in which gravity plays an important part in keeping the wormhole throat open, removing the necessity for exotic matter. Kuhfittig \cite{P. K. H. Kuhfittig} has studied rotating axially symmetric wormholes by introducing time-dependent angular velocity to generalize static and spherically symmetric traversable wormholes. B\"ohmer et al. \cite{C. G. Bohmer 1} investigated wormhole solutions by assuming a linear relationship between energy density and pressure. Barros and Lobo \cite{B. J. Barros} developed wormhole solutions using three-form fields and explored their compatibility with weak and null energy conditions. Also, stability of thin shell wormholes with barotropic fluids present has been done in \cite{N. Tsukamoto}. Wormhole geometries have garnered considerable attention not only in modified gravity theories but also in higher-dimensional gravitational theories \cite{Mak, Zangeneh, Galiakhmetov, Kar2, Ziaie} and Kaluza-Klein gravity \cite{Singleton, Leon, Folomeev}. These theoretical frameworks offer distinct advantages, such as eliminating the need for nonstandard fluids, which has been a significant driving force behind extensive research in modified gravity theories. Moreover, modifications to Einstein's gravity introduced additional degrees of freedom within the gravitational sector, presenting new avenues for addressing challenges such as dark energy and dark matter. Additionally, Falco et al. \cite{V. De Falco} concentrated on $f(R)$ metric, $f(Q)$ symmetric teleparallel, and $f(T)$ teleparallel models to explore spherically symmetric static wormhole solutions. Recent literature \cite{Karakasis, Golchin, Eid, Goswami, Malik, Ahmad, Bhatti, Chanda, Rosa, Tayde 1, Tayde 2, C. G. Bohmer 2, Rani, Momeni, Tayde 3, Pavlovic} provides an insightful investigation of wormhole geometry under several modified gravity theories. These works contribute substantially to the ongoing debate about the theoretical underpinnings and observational implications associated with wormholes in alternative gravitational contexts.\\
Jimenez et al. \cite{J. B. Jimenez} proposed $f(Q)$ gravity, which relies on the nonmetricity $Q$ to explain gravitational interactions. Over the past few years, $f(Q)$ gravity has been under strong observational investigation, with Lazkoz et al. \cite{R. Lazkoz} suggesting significant restrictions on its validity. Their study examined the plausibility of $f(Q)$ gravity using a variety of observable datasets, including the Type Ia Supernovae, expansion rate, quasars, baryon acoustic oscillations, gamma-ray bursts, and cosmic microwave background distance. Mandal et al. \cite{S. Mandal} established the validity of $f(Q)$ gravity models under energy conditions. They developed an embedding approach that incorporates the non-trivial contributions of the nonmetricity function into energy restrictions, confirming the consistency of $f(Q)$ gravity with fundamental physical laws. These efforts contribute substantially to understanding the potential of $f(Q)$ gravity as a convincing alternative to traditional gravitational theories, clearing the way for further investigation into the basic concept of gravity and its significance for cosmology. Also, Gadbail et al. \cite{G. Gadbail} presented $f(Q)$ gravity reconstruction from Friedmann-Laimatre-Robertson-Walker (FLRW) evolution, demonstrating exact $\Lambda$CDM expansion and showing how generic functions of $Q$ require extra degrees of freedom in the matter part. Furthermore, the impact of the Generalized Uncertainty Principle (GUP) on the Casimir wormhole space-time in $f(Q)$ gravity \cite{Zinnat 2} has been analyzed systematically. In their investigation, the authors considered two distinct forms of $f(Q)$ and meticulously derived both analytic and numerical solutions. These solutions were obtained by incorporating the influence of the GUP. Besides that, they investigated the ADM mass and used volume integral quantifiers to calculate the required amount of exotic matter at the throat of the wormhole. Also, Hassan et al. \cite{Zinnat 3} studied how the Casimir effect affects wormhole geometry under $f(Q)$ gravity. To improve experimental feasibility, they looked at three Casimir effect systems: (i) two parallel plates, (ii) two parallel cylindrical plates, and (iii) two spheres separated by a significant distance and discovered that certain arbitrary quantities violated classical energy conditions at the wormhole throat. Importantly, various intriguing astrophysical experiments have been carried out using the $f(Q)$ gravity theory and its extended gravity. References \cite{L. Heisenberg, F. Parsaei, O. Sokoliuk 1, O. Sokoliuk 3, P. Bhar, S. Pradhan, Debasmita, Tayde 5} provide valuable resources for further investigation.\\
\indent Monopoles were first proposed by Dirac \cite{P. A. M. Dirac} when he had shown by geometric arguments that if magnetic monopoles exist, then the charge of monopoles would be quantized. Even though the electromagnetic field is a rather simple abelian (U(1)) gauge field, it was later shown by G 't Hooft \cite{G.'t Hooft} and Polykov \cite{A. M. Polyakov} that more complicated non-abelian gauge fields would also lead to similar types of monopoles. As we know, the hot Big Bang model of the universe in the Plank era had all the forces in equilibrium, and then the four forces decoupled; this gives a rather compelling argument that, indeed, we can find the monopoles from the early universe, which was produced by the spontaneous symmetry breaking. In this article, we rather focused on a simple Barriola valentine \cite{M. Barriola} type monopole (which forms when $SO(3)$ group spontaneous symmetry breaks into $SO(2)$ group) for which the effect of the gravitational field is well known. Moreover, Zloshchastiev \cite{K. G. Zloshchastiev} gave a comparative description of monopole in the quantum mechanical context. Also, De-Chang Da et al. \cite{De Chang Da} have studied the phenomenological aspects of the wormhole under the influence of global monopole. Further,  Ahmed \cite{F. Ahmed} studied a topologically charged four-dimensional wormhole and the energy conditions, including global monopole charge, and found that the energy-momentum tensor associated with this wormhole complies with the weak energy condition and the null energy condition. In reference \cite{X. Shi}, Shi et al. provided an exact solution for the nonlinear equation that describes the global monopole (GM) in flat space. Their study of the metric outside the GM revealed a repulsive gravitational field beyond the core and a solid angular deficit. Reference \cite{D. P. Bennett} demonstrated that the gravitational fields of GMs can cause matter clustering and anisotropies in the Cosmic Microwave Background (CMB). Additionally, reference \cite{R. H. Brandenberger} suggested that GMs might contribute to forming supermassive black holes. Studies \cite{X. Shi, D. Harari} have shown that this topological defect induces a negative gravitational potential, leading to a repulsive gravitational field. Many researchers have highlighted the essential role of topological defects in the formation of cosmological structures and the evolution of the cosmos, as noted in references \cite{M. Yamaguchi, A. Vilenkin, R. Basu, R. Durrer1, R. Durrer2}. Recently, reference \cite{M. Kalam} investigated wormholes within the Milky Way galaxy that possess a global monopole charge. This article investigated whether such a monopole could cause the wormhole in modified $f(Q)$ gravity. It is well known that wormholes can be sustained by global monopoles. Here, we have investigated whether the same holds in symmetric teleparallel gravity. While previous studies have examined wormholes in the context of monopoles \cite{Rahaman11, S. Sarkar, P. Das, K. Jususfi}, we provide a more comprehensive perspective by considering $f(Q)$ gravity.\\
\indent The organization of this paper is as follows: Section \ref{sec2} outlines the criteria defining a traversable wormhole imbued with a Global Monopole and presents the formalism elucidating $f(Q)$ gravity. In Section \ref{sec3}, we derive the field equations within a linear model framework. Subsequently, in Section \ref{sec4}, three linear EoS cases are introduced, and the necessary conditions for the existence of a traversable wormhole are examined, along with a scrutiny of the energy conditions under three different choices of redshift function. Section \ref{sec5} encompasses a stability analysis conducted utilizing the Tolman-Oppenheimer-Volkov (TOV) equation. Finally, Section \ref{sec6} consolidates the findings and engages in a comprehensive discussion concerning the implications of the results derived from this study.

\section{Basic Criteria of a traversable wormhole with Global Monopole and formalism of $f(Q)$ gravity}
\label{sec2}
The wormhole metric in the Schwarzschild coordinates $(t,r,\theta,\Phi)$, as defined by \cite{M. Visser 1, M. S. Morris}, is expressed as:
\begin{equation}\label{11}
\hspace{-0.2cm}ds^2=e^{2\phi(r)}dt^2-\left(1-\frac{b(r)}{r}\right)^{-1}dr^2-r^2\,d\theta^2-r^2\,\sin^2\theta\,d\Phi^2\,.
\end{equation}
Here, $b(r)$ denotes the shape function defining the geometry of the wormhole. The function $\phi(r)$ shows the redshift function associated with gravitational redshift. To render a wormhole traversable, the shape function $b(r)$ must satisfy the flaring-out condition, as specified by $(b-b'r)/b^2>0$ \cite{M. S. Morris}. At the wormhole throat $b(r_0)=r_0$, the condition $b^{\,\prime}(r_0)<1$ is imposed, where $r_0$ denotes the throat radius. Additionally, the asymptotic flatness condition demands that as $r\rightarrow \infty$, the ratio $\frac{b(r)}{r}$ tends to $0$. Furthermore, to avoid an event horizon, $\phi(r)$ must remain finite at all points. Adherence to these conditions ensures the possibility of exotic matter existence at the wormhole throat within the framework of Einstein's GR.

Now, the formulation of the action for symmetric teleparallel gravity with a global monopole charge, derived from a four-dimensional action without considering the cosmological constant and minimally coupled triplet scalar field ($\hbox {c}=\hbox {G}=1$), can be expressed as follows:
\begin{equation}\label{1}
\mathcal{S}=\int\frac{1}{16\pi}\,f(Q)\sqrt{-g}\,d^4x+\int (\mathcal{L}_m+\mathcal{L})\,\sqrt{-g}\,d^4x\,.
\end{equation}
In the given context, $f(Q)$ denotes an arbitrary function of $Q$, where $Q$ represents the non-metricity scalar. $\mathcal{L}_m$ signifies the matter Lagrangian density, $\mathcal{L}$ denotes the Lagrangian density of the monopole, and $g$ stands for the determinant of the metric tensor $g_{\mu\nu}$.\\
The Lagrangian density of a self-coupling scalar triplet $\phi ^a$ is expressed as:
\begin{equation}\label{1a}
{\mathcal {L}}= - \frac{\lambda }{4}(\phi ^2 -\eta ^2)^2 -\frac{1}{2} \sum _a g^{ij} \partial _i \phi ^a \partial _j \phi ^a.
\end{equation}
In this expression, $a=1,2,3$, $\eta$, and $\lambda$ represent the gauge-symmetry breaking and self-interaction term scales, respectively. The field configuration corresponding to the monopole is described as follows:
\begin{equation}\label{1b}
\phi ^a = \frac{\eta }{r} F(r) x^a\,,
\end{equation}
where the variable $x^a$ is defined as $(r sin\theta cos\phi , r sin\theta sin\phi , rcos\theta )$, ensuring that $\sum _a x^a x^2 = r^2$.\\
The Lagrangian density can be expressed in terms of $F(r)$ by utilizing the field configuration as
\begin{equation}\label{1c}
\mathcal{L} = -\left( 1-\frac{b(r)}{r}\right) \frac{\eta ^2 (F')^2}{2} -\frac{\eta ^2 F^2}{r^2}-\frac{\lambda \eta ^4}{4}(F^2-1)^2,
\end{equation}
and for the field $F(r)$, the Euler-Lagrangian equation is
\begin{multline}\label{1d}
\left( 1-\frac{b(r)}{r}\right) F'' + F' \left[ \left( 1-\frac{b(r)}{r}\right) \frac{2}{r}+\frac{1}{2}\left( \frac{b-b' r}{r^2}\right) \right] \\
-F\left[ \frac{2}{r^2}+\lambda \eta ^2(F^2-1)\right] =0 .
\end{multline}
The energy-momentum tensor is constructed from Eq. \eqref{1a} as
\begin{multline}\label{1e}
\bar{T}_{ij}=\partial _i\phi ^a\partial _j\phi ^a -\frac{1}{2} g_{ij}g^{\mu \nu }\partial _\mu \phi ^a\partial _\nu \phi ^a -\frac{g_{ij}\lambda }{4}(\phi ^2 -\eta ^2)^2.
\end{multline}
Hence, all four elements of the Energy-Momentum tensor can be found using Eq. \eqref{1e}:
\begin{equation}\label{1f}
\bar{T}^t_ t=-\eta ^2\left[ \frac{F^2}{r^2}+\left( 1-\frac{b(r)}{r}\right) \frac{(F')^2}{2} +\frac{\lambda \eta ^2}{4}(F^2-1)^2\right],
\end{equation}
\begin{equation}\label{1g}
\bar{T}^r_ r=-\eta ^2\left[ \frac{F^2}{r^2}+\left( 1-\frac{b(r)}{r}\right) \frac{(F')^2}{2} +\frac{\lambda \eta ^2}{4}(F^2-1)^2\right],
\end{equation}
\begin{multline}\label{1h}
\bar{T}^\theta _ \theta=\bar{T}^\Phi _\Phi \nonumber =-\eta ^2\left[ \left( 1-\frac{b(r)}{r}\right) \frac{(F')^2}{2} +\frac{\lambda \eta ^2}{4}(F^2-1)^2\right].
\end{multline}
As demonstrated in Eq. \eqref{1d}, obtaining a precise analytical solution can be challenging. Therefore, it is often advantageous to approximate the region outside the wormhole to simplify the analysis and derive results. As a result, the components of the reduced energy-momentum can be expressed as follows:
\begin{equation}\label{1i}
\bar{T}^t_t=\bar{T}^r_r=-\frac{\eta ^2}{r^2} , \bar{T}^\theta _\theta =\bar{T}^\Phi _\Phi = 0.
\end{equation}
Furthermore, the non-metricity tensor can be expressed by the following equation \cite{J. B. Jimenez}.
\begin{equation}\label{2}
Q_{\lambda\mu\nu}=\bigtriangledown_{\lambda} g_{\mu\nu}\,.
\end{equation}
Additionally, the superpotential, also known as the non-metricity conjugate, has the following formal definition:
\begin{equation}\label{3}
\hspace{-0.2cm}P^\alpha\,_{\mu\nu}=\frac{1}{4}\left[-Q^\alpha\;_{\mu\nu}+2Q_{(\mu}\;^\alpha\;_{\nu)}+Q^\alpha g_{\mu\nu}-\tilde{Q}^\alpha g_{\mu\nu}-\delta^\alpha_{(\mu}Q_{\nu)}\right].
\end{equation}
Also, traces of the non-metricity tensor can be provided by
\begin{equation}
\label{4}
\tilde{Q}_\alpha=Q^\mu\;_{\alpha\mu}\,,\;Q_{\alpha}=Q_{\alpha}\;^{\mu}\;_{\mu}.
\end{equation}
The non-metricity scalar is given by \cite{J. B. Jimenez}
\begin{eqnarray}
\label{5}
Q &=& -P^{\alpha\mu\nu}\,Q_{\alpha\mu\nu}\\
&=& g^{\mu\nu}\left(L^\beta_{\,\,\,\alpha\beta}\,L^\alpha_{\,\,\,\mu\nu}-L^\beta_{\,\,\,\alpha\mu}\,L^\alpha_{\,\,\,\nu\beta}\right),
\end{eqnarray}
The disformation tensor, denoted as $L^\beta_{\,\,\,\mu\nu}$, is defined as follows:
\begin{equation}\label{6}
L^\beta_{\,\,\,\mu\nu}=\frac{1}{2}Q^\beta_{\,\,\,\mu\nu}-Q_{(\mu\,\,\,\,\,\,\nu)}^{\,\,\,\,\,\,\beta}.
\end{equation}
The equations of motion for gravity can be obtained by varying the action concerning the metric tensor $g_{\mu\nu}$. This equation is expressed as follows:
\begin{multline}\label{7}
\frac{-2}{\sqrt{-g}}\bigtriangledown_\alpha\left(\sqrt{-g}\,f_Q\,P^\alpha\;_{\mu\nu}\right)-\frac{1}{2}g_{\mu\nu}f \\
-f_Q\left(P_{\mu\alpha\beta}\,Q_\nu\;^{\alpha\beta}-2\,Q^
{\alpha\beta}\,\,_{\mu}\,P_{\alpha\beta\nu}\right)=8\pi T_{\mu\nu}\,,
\end{multline}
where $f_Q=\frac{\partial f}{\partial Q}$ and $T_{\mu\nu}$ is the sum of the energy-momentum tensor of anisotropic fluid part and matter field part. Therefore, it can be written as
\begin{equation}\label{8}
T_{\mu \nu } =\mathcal{T}_{\mu \nu } + \bar{T}_{\mu \nu }\,.
\end{equation}
The elements of the energy-momentum for the anisotropic fluid are
\begin{equation}\label{9}
\mathcal{T}^\mu\;_\nu =\text{diag}(\rho,-p_r,-p_t,-p_t),
\end{equation}
where $\rho$ is the energy density, $p_r$ is the radial pressure, and $p_t$ is tangential pressure.
The non-metricity scalar $Q$, associated with metric \eqref{11}, is defined in accordance with the reference \cite{Tayde 4} as
\begin{equation}\label{10}
Q=-\frac{b}{r^2}\left[2\phi^{'}+\frac{rb^{'}-b}{r(r-b)}\right].
\end{equation}\\
Therefore, the field equations governing $f(Q)$ gravity for the wormhole, taking into account the Global Monopole Charge, can be expressed as:
\begin{multline}\label{12}
8 \pi  \rho =\frac{(r-b)}{2 r^3} \left[f_Q \left(\frac{(2 r-b) \left(r b'-b\right)}{(r-b)^2}+\frac{b \left(2 r \phi '+2\right)}{r-b}\right)
\right. \\ \left.
+\frac{f r^3}{r-b}+\frac{2 b r f_{\text{QQ}} Q'}{r-b}\right]-\frac{8 \pi  \eta ^2}{r^2},
\end{multline}
\begin{multline}\label{13}
8 \pi  p_r=-\frac{(r-b)}{2 r^3} \left[-f_Q \left(\frac{b }{r-b}\left(\frac{r b'-b}{r-b}+2+2 r \phi '\right)
\right.\right. \\ \left.\left.
-4 r \phi '\right)-\frac{f r^3}{r-b}-\frac{2 b r f_{\text{QQ}} Q'}{r-b}\right]+\frac{8 \pi  \eta ^2}{r^2},
\end{multline}
\begin{multline}\label{14}
8 \pi  p_t=\frac{(r-b)}{4 r^2} \left[f_Q \left(\frac{\left(r b'-b\right) \left(\frac{2 r}{r-b}+2 r \phi '\right)}{r (r-b)}+
\right.\right. \\ \left.\left.
\frac{4 (2 b-r) \phi '}{r-b}-4 r \left(\phi '\right)^2-4 r \phi ''\right)+\frac{2 f r^2}{r-b}
\right.\\\left.
-4 r f_{\text{QQ}} Q' \phi '\right].
\end{multline}
Utilizing these precise field equations, a comprehensive exploration of different wormhole solutions within the framework of $f(Q)$ gravity models becomes feasible.

\subsection{Energy conditions}
The classical energy conditions, stemming from the Raychaudhuri equations, constitute a fundamental framework for exploring physically feasible matter configurations. These conditions, which encompass weak, null, strong, and dominant categories, play a crucial role in understanding the properties of matter within the context of GR. Among them, the null energy condition assumes particular significance, especially in the context of GR's treatment of wormhole solutions. The null energy condition holds pivotal importance due to its direct relationship with the energy density necessary to maintain the integrity of a wormhole's throat. Any departure from the null energy condition near the throat of a wormhole implies the existence of exotic matter characterized by negative energy density- an attribute not typically associated with conventional matter sources. Energy conditions function as constraints on the stress-energy tensor, delineating the distribution of matter and energy in space-time. They are represented as follows:\\
$\bullet$ The weak energy condition (\textbf{WEC}) :
$\rho\geq0$,\,\, $\rho+p_t\geq0$,\,\, and \,\, $\rho+p_r\geq0$.\\
$\bullet$ The null energy condition (\textbf{NEC}) : $\rho+p_t\geq0$\,\, and \,\, $\rho+p_r\geq0$.\\
$\bullet$ The dominant energy condition (\textbf{\textbf{DEC}}) :  $\rho\geq0$,\,\, $\rho+p_t\geq0$,\,\, $\rho+p_r\geq0$,\,\, $\rho-p_t\geq0$,\,\, and \,\, $\rho-p_r\geq0$.\\
$\bullet$ The strong energy condition (\textbf{SEC}) :
 $\rho+p_t\geq0$,\,\, $\rho+p_r\geq0$,\,\, and \,\, $\rho+p_r+2p_t\geq0$.\\
In conclusion, energy conditions provide crucial constraints on the behavior of matter in the Universe and play a crucial role in our exploration of wormholes.
\section{Linear $f(Q)$ model}\label{sec3}
In the present study, we adopt a linear functional form of $f(Q)$ gravity, represented as:
\begin{equation}\label{15}
f(Q)=\alpha Q+\beta\,.
\end{equation}
Here, $\alpha\neq0$ represents a model parameter. This value can easily be modified to regain GR when $\alpha=1$ and $\beta=0$. Using this linear model, Solanki et al. \cite{R solanki} explored the late-time cosmic acceleration without including any dark energy component in the matter part. Avik De and Tee-How Loo's work \cite{Avik De} shows that this model maintains energy conservation. This model is used in \cite{G. Mustafa} to analyze traversable wormholes inspired by non-commutative geometries with conformal symmetry, indicating the possibility of wormhole solutions with viable physical features. Moreover, this symmetric teleparallel model is also used in Casimir wormhole \cite{Zinnat 3, Zinnat 2}. Also, this model successfully explains strange stars, which is consistent with observation \cite{S V Lohakare}. Further, S. K. Maurya et al. \cite{S. K. Maurya} attempted to find an anisotropic solution for a compact star generated by this linear form of the $f(Q)$ model. One can look into the following literature \cite{A. Ditta, A. Errehymy} for more extensive studies conducted in the context of symmetric teleparallel gravity. Therefore, the reduced field equations with an arbitrary redshift function $\phi(r)$ are stated as follows:
\begin{equation}\label{16}
\rho=\frac{\alpha  b'-8 \pi  \eta ^2}{8 \pi  r^2},
\end{equation}
\begin{equation}\label{17}
p_r = \frac{2 \alpha  r (b-r) \phi '+\alpha  b+8 \pi  \eta ^2 r}{8 \pi  r^3},
\end{equation}
\begin{equation}\label{18}
p_t = \frac{\alpha  \left(r \phi '+1\right) \left(r b'+2 r (b-r) \phi '-b\right)+2 \alpha  r^2 (b-r) \phi ''}{16 \pi  r^3}.
\end{equation}
This study introduces three different forms of redshift function $\phi(r)$ as a component in deriving the wormhole solution. We introduce additional cases and study the energy conditions to obtain analytical solutions for the resulting field equations. These will be elaborated in the next Section \ref{sec4}.

\section{Linear equation of states}\label{sec4}
In this particular section, we consider three different EoS such as barotropic ($p_r=\omega \rho$), anisotropic ($p_t=n p_r$), and isotropic ($p_t=p_r$) to find the exact analytical solution ($b(r)$). These cases are as follows:
\subsection{Barotropic EoS}\label{subsec1}
In this section, we are delving into the construction of wormhole solutions by utilizing a linear barotropic Equation of State (EoS) as presented in \cite{F. S. N. Lobo 1}:
\begin{equation}\label{19}
p_r=\omega \rho   \,.
\end{equation}
Here, $\omega$ represents the EoS parameter. In late-time cosmic acceleration, the description of dark energy within the $\Lambda$CDM framework is distinguished by an EoS parameter of $\omega=-1$, which has exhibited remarkable success thus far. Another extensively discussed model for time-dependent dark energy is the quintessence model, characterized by an EoS parameter $\omega>-1$. Conversely, the least theoretically understood form of dark energy is phantom energy, identified by an EoS parameter $\omega<-1$. Reference \cite{F. S. N. Lobo 2} explores asymptotically flat phantom wormhole solutions. Additionally, references \cite{Z. Hassan 1, K. N. Singh} point out the considerable challenge of discovering wormhole solutions with a linear barotropic EoS within the context of non-linear models in teleparallel and symmetric teleparallel gravity. Also, Solanki et al. \cite{R. Solanki} successfully derived the precise wormhole solution within the framework of the non-linear $f(R, L_m)$ model while maintaining a linear barotropic Equation of State (EoS). Nevertheless, we manage to derive an exact wormhole solution within the framework of our linear $f(Q)$ model while adhering to a linear barotropic EoS using redshift functions as discussed in below Subsubsections \ref{subsubsec1}-\ref{subsubsec3}.\\

\subsubsection{$\phi(r)=c$}\label{subsubsec1}
\begin{figure*}[t]
    \centering
    \includegraphics[width=17.5cm,height=5cm]{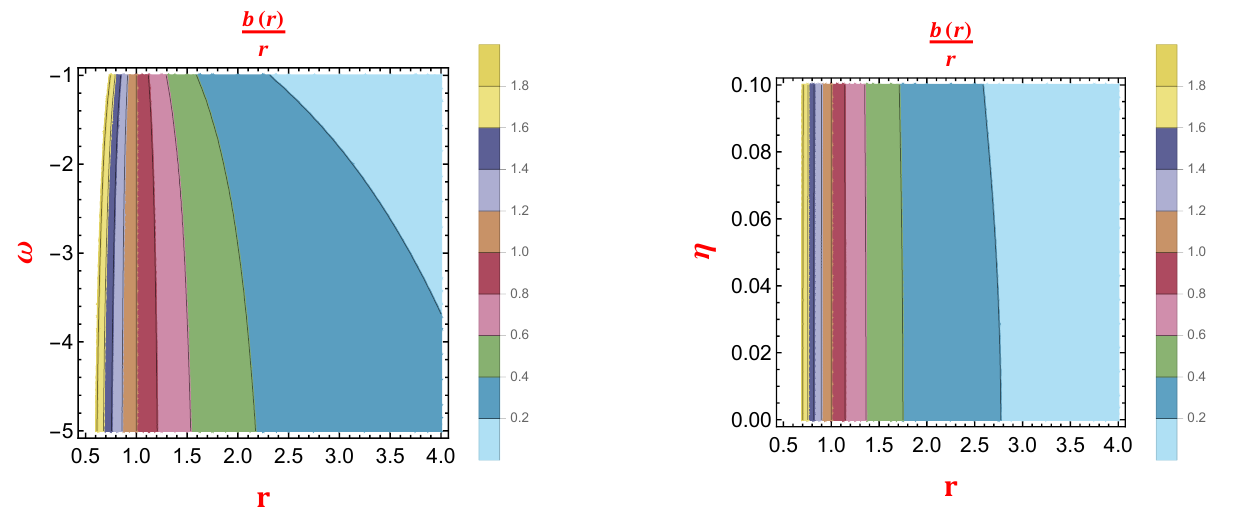}
    \caption{The contour plot illustrates the barotropic EoS, depicting changes in the asymptotic flatness condition for $\omega$ \textit{(depicted on the left)} and $\eta$ \textit{(depicted on the right)} concerning the radial coordinate `$r$'. Additionally, the plot maintains constant values for other parameters, such as $\alpha=-2$, $\beta=0.02$, and $r_0 = 1$.}
    \label{fig1}
\end{figure*}

\begin{figure*}[t]
    \centering
    \includegraphics[width=17.5cm,height=5cm]{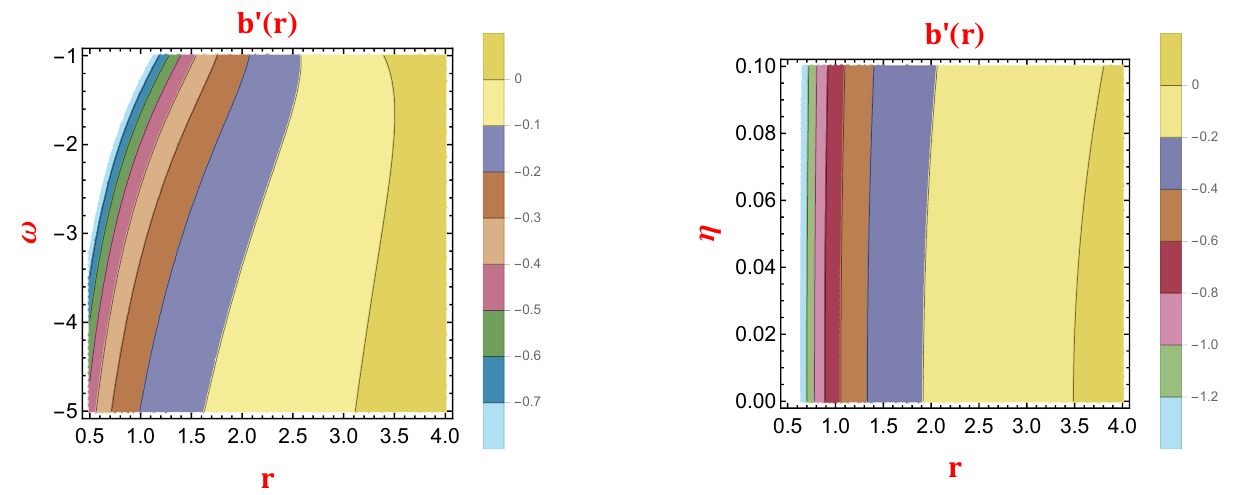}
    \caption{The contour plot illustrates the barotropic EoS, depicting changes in the flaring-out condition for $\omega$ \textit{(depicted on the left)} and $\eta$ \textit{(depicted on the right)} concerning the radial coordinate `$r$'. Additionally, the plot maintains constant values for other parameters, such as $\alpha=-2$, $\beta=0.02$, and $r_0 = 1$.}
    \label{fig2}
\end{figure*}

\begin{figure*}[t]
\centering
\includegraphics[width=14.5cm,height=5cm]{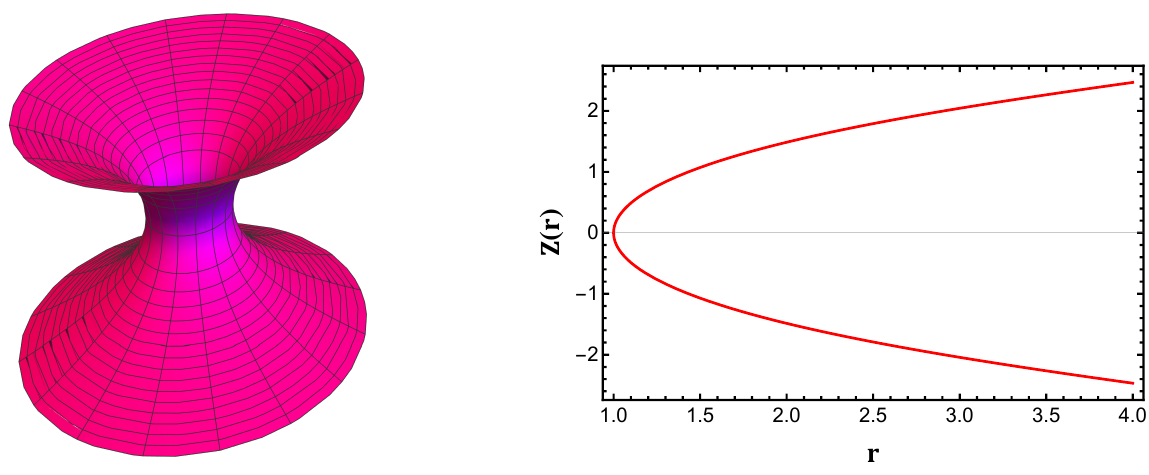}
\caption{The figure displays the embedding diagram for the barotropic EoS. Additionally, the figure maintains constant values for other parameters, such as $\alpha=-2$, $\beta=0.02$, $\eta=0.02$, $\omega=-1.5$ and $r_0 = 1$.}
\label{fig13}
\end{figure*}

\begin{figure*}[t]
    \centering
    \includegraphics[width=17.5cm,height=5cm]{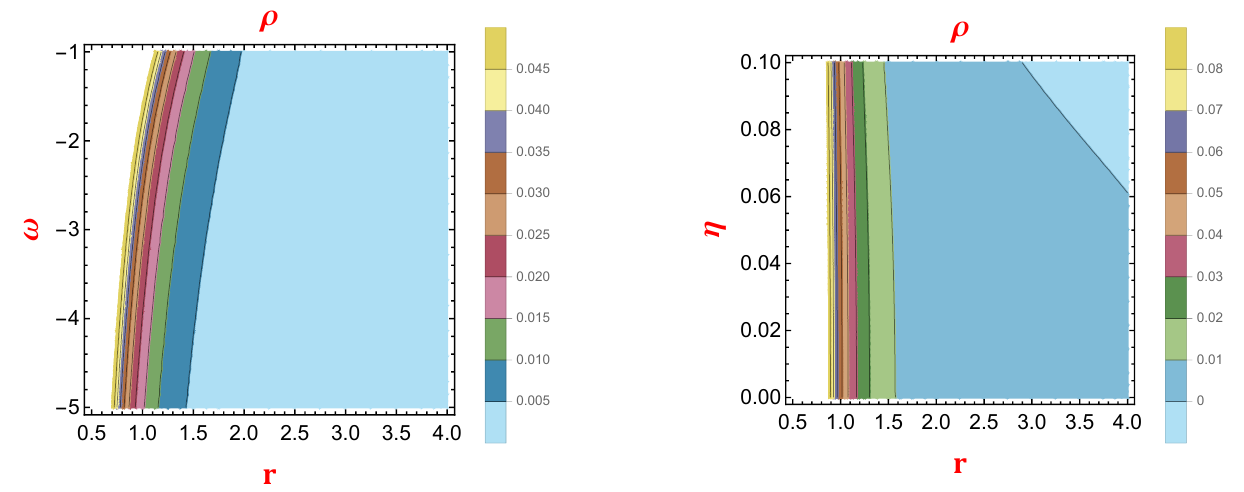}
    \caption{The contour plot illustrates the barotropic EoS, depicting changes in the energy density for $\omega$ \textit{(depicted on the left)} and $\eta$ \textit{(depicted on the right)} concerning the radial coordinate `$r$'. Additionally, the plot maintains constant values for other parameters, such as $\alpha=-2$, $\beta=0.02$, and $r_0 = 1$.}
    \label{fig3}
\end{figure*}

\begin{figure*}[t]
    \centering
    \includegraphics[width=17.5cm,height=5cm]{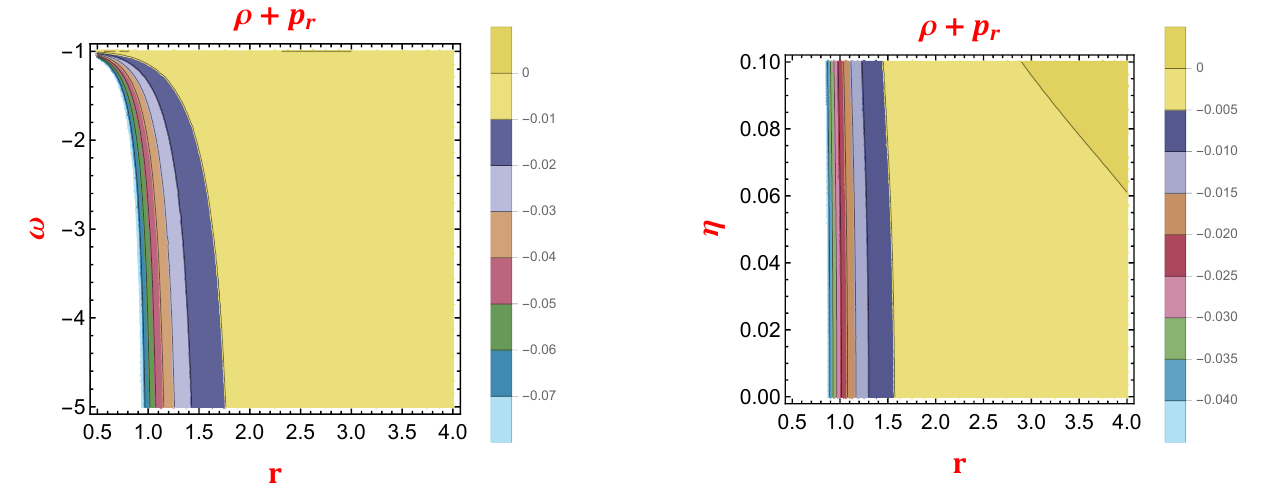}
    \caption{The contour plot illustrates the barotropic EoS, depicting changes in the radial pressure for NEC for $\omega$ \textit{(depicted on the left)} and $\eta$ \textit{(depicted on the right)} concerning the radial coordinate `$r$'. Additionally, the plot maintains constant values for other parameters, such as $\alpha=-2$, $\beta=0.02$, and $r_0 = 1$.}
    \label{fig4}
\end{figure*}

\begin{figure*}[t]
    \centering
    \includegraphics[width=17.5cm,height=5cm]{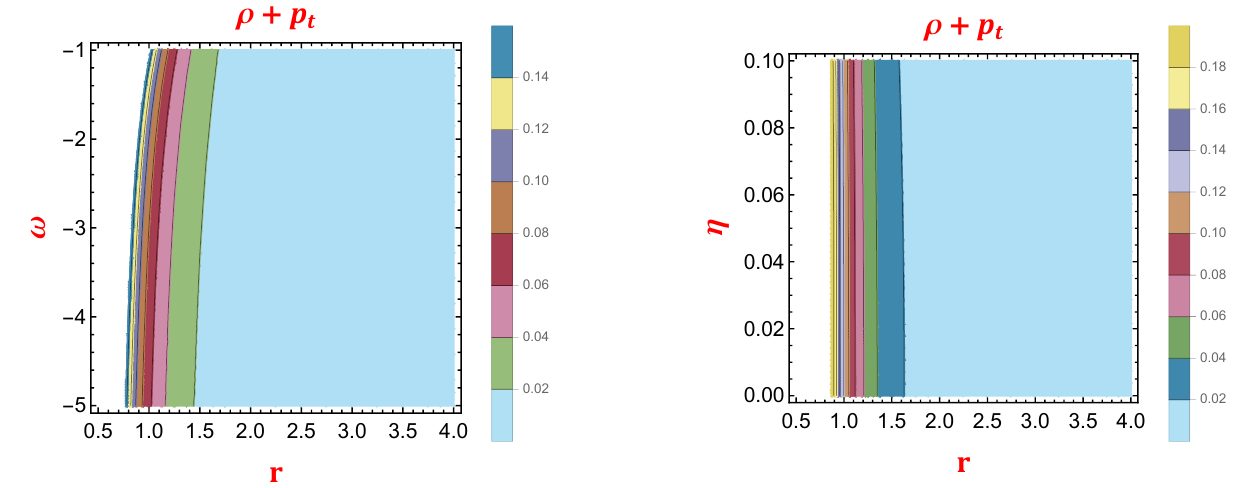}
    \caption{The contour plot illustrates the barotropic EoS, depicting changes in the tangential pressure for NEC for $\omega$ \textit{(depicted on the left)} and $\eta$ \textit{(depicted on the right)} concerning the radial coordinate `$r$'. Additionally, the plot maintains constant values for other parameters, such as $\alpha=-2$, $\beta=0.02$, and $r_0 = 1$.}
    \label{fig5}
\end{figure*}

\begin{figure*}[t]
    \centering
    \includegraphics[width=17.5cm,height=5cm]{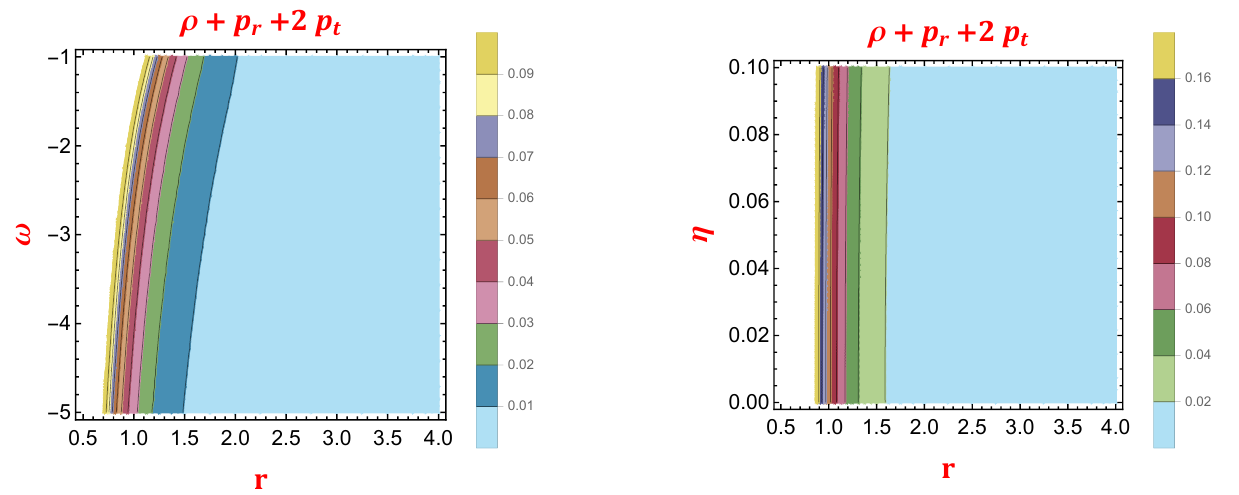}
    \caption{The contour plot illustrates the barotropic EoS, depicting changes in the SEC for $\omega$ \textit{(depicted on the left)} and $\eta$ \textit{(depicted on the right)} concerning the radial coordinate `$r$'. Additionally, the plot maintains constant values for other parameters, such as $\alpha=-2$, $\beta=0.02$, and $r_0 = 1$.}
    \label{fig6}
\end{figure*}

The function $\phi(r) = c$ \cite{Z. Hassan 1, N. Godani}, where $c$ represents a constant, is widely regarded as an appropriate selection for a redshift function owing to its consistent behavior with respect to the radial coordinate $r$. Notably, when $c$ equals zero, this function is commonly referred to as the tidal force. Upon solving equations \eqref{16} and \eqref{17} while incorporating the relation \eqref{19}, we arrive at the following first-order differential equation given by
\begin{multline}\label{20}
b'(r)-\left(\frac{1}{r \omega }\right)b(r)=\frac{16 \pi  \eta ^2 (\omega +1)}{2 \alpha  \omega }-\frac{\beta  r^2 (\omega -1)}{2 \alpha  \omega }\,.
\end{multline}
Upon integrating the aforementioned equation with the throat condition $b(r_0)=r_0$, we get the explicit form of shape function as
\begin{multline}\label{21}
\hspace{-0.2cm}b(r)=r_0^{-1/\omega } r^{1/\omega } \left(\frac{r_0^3 \beta  (\omega -1)}{2 \alpha  (3 \omega -1)}-\frac{8 \pi  r_0 \eta ^2 (\omega +1)}{\alpha  (\omega -1)}+r_0\right)  \\
+\frac{8 \pi  \eta ^2 r (\omega +1)}{\alpha  (\omega -1)}-\frac{\beta  r^3 (\omega -1)}{2 \alpha  (3 \omega -1)}\,.
\end{multline}
Now, we will delve into the visual representation of the shape function and explore the crucial conditions for the existence of a wormhole. To accomplish this, we will meticulously select appropriate parameters. We will showcase the graphical depiction of the asymptotic behavior of the shape function in Contour plot \ref{fig1} for the EoS parameter $\omega$ (in the left plot) and global monopole parameter $\eta$ (in the right plot). It confirms that as the radial distance increases, the ratio $\frac{b(r)}{r}$ tends towards $0$. In addition, we verify the satisfaction of the flaring-out condition $b'(r_0)<1$ at the wormhole throat $r=r_0$ through Contour plot \ref{fig2} for the same parameters $\omega$ and $\eta$. Here, we designate the wormhole throat at $r_0=1$. Also, we present the embedding diagram and its comprehensive visualization in Fig. \ref{fig13}.\\
By employing equation \eqref{21} in equations \eqref{16}-\eqref{18}, one can obtain the expressions for the energy density, radial pressure, and tangential pressure as follows
\begin{multline}\label{22}
\rho= \frac{1}{16 \pi  r^2} \left(2\alpha \left(\frac{8 \pi  \eta ^2 (\omega +1)}{\alpha  (\omega -1)}-\frac{3\beta  r^2 (\omega -1)}{2 \alpha  (3 \omega -1)}
\right.\right. \\ \left.\left.
+\frac{r_0^{-1/\omega } r^{\frac{1}{\omega }-1}}{\omega} \left(\frac{r_0^3 \beta  (\omega -1)}{2 \alpha  (3 \omega -1)}-\frac{8 \pi  r_0 \eta ^2 (\omega +1)}{\alpha  (\omega -1)}+r_0\right)\right)
\right. \\ \left.
+\beta  r^2-16 \pi  \eta ^2\right)\,,
\end{multline}
\begin{multline}\label{23}
p_r= \frac{1}{16 \pi  r^3} \left(2\alpha \left(\frac{8 \pi  \eta ^2 r (\omega +1)}{\alpha  (\omega -1)}-\frac{\beta  r^3 (\omega -1)}{2 \alpha  (3 \omega -1)}
\right.\right. \\ \left.\left.
+r_0^{-1/\omega } r^{1/\omega } \left(\frac{r_0^3 \beta  (\omega -1)}{2 \alpha  (3 \omega -1)}-\frac{8 \pi  r_0 \eta ^2 (\omega +1)}{\alpha  (\omega -1)}+r_0\right)\right)
\right. \\ \left.
+\beta  r^3+16 \pi  \eta ^2 r\right)\,,
\end{multline}
\begin{multline}\label{24}
p_t= \frac{1}{16 \pi  r^3}\left(\alpha \left(\frac{-8 \pi  \eta ^2 r (\omega +1)}{\alpha  (\omega -1)}+\frac{\beta  r^3 (\omega -1)}{2 \alpha  (3 \omega -1)}
\right.\right. \\ \left.\left.
-r_0^{-1/\omega } r^{1/\omega } \left(-\frac{8 \pi  r_0 \eta ^2 (\omega +1)}{\alpha  (\omega -1)}+r_0+\frac{r_0^3 \beta  (\omega -1)}{2 \alpha  (3 \omega -1)}\right) 
\right.\right. \\ \left.\left.
+ r\left(\frac{8 \pi  \eta ^2 (\omega +1)}{\alpha  (\omega -1)}-\frac{3 \beta  r^2 (\omega -1)}{2 \alpha  (3 \omega -1)}+\frac{r_0^{-1/\omega } r^{\frac{1}{\omega }-1}}{\omega}
\right.\right.\right. \\ \left.\left.\left.
\times \left(\frac{r_0^3 \beta  (\omega -1)}{2 \alpha  (3 \omega -1)}-\frac{8 \pi  r_0 \eta ^2 (\omega +1)}{\alpha  (\omega -1)}+r_0\right)\right)\right)+\beta  r^3\right)\,.
\end{multline}
Now, we present the graphical representation of the energy density using Contour plot \ref{fig3}, illustrating its positively decreasing behavior throughout the Space-time for the parameters $\omega$ and $\eta$ concerning the radial coordinate $r$. Additionally, NEC at the throat of wormhole $r=r_0$ is given by
\begin{equation}
\left(\rho + p_r\right)_{\text{at}\,\, r=r_0}=\frac{(\omega +1) \left(r_0^2 \beta +2 \alpha +16 \pi  \eta ^2\right)}{16 \pi  r_0^2 \omega }\,,
\end{equation}
\begin{equation}
\left(\rho + p_t\right)_{\text{at}\,\, r=r_0}=\frac{(\omega +3) \left(r_0^2 \beta +16 \pi  \eta ^2\right)-2 \alpha  (\omega -3)}{32 \pi  r_0^2 \omega }\,.
\end{equation}
Moreover, we present the graphical illustration of this condition in Contour plots \ref{fig4} and \ref{fig5}, providing a visual understanding of the radial and tangential pressures at the wormhole throat. Through Contour plot \ref{fig4}, it is clear that NEC for radial pressure shows negatively increasing behavior as radial distance increases. Also, the Contour plot \ref{fig5} shows the positive behavior of NEC for tangential pressure. Overall, NEC is violated in this barotropic case, which may show the presence of exotic matter that maintains stability at the wormhole's throat. In addition, we extend our analysis toward SEC, where we observe satisfaction near the throat for both parameters and can be visualized through Contour plot \ref{fig6}.

\subsubsection{$\phi(r)=\log\left(1+\frac{r_0}{r}\right)$}\label{subsubsec2}
\begin{figure*}[t]
    \centering
    \includegraphics[width=17.5cm,height=5cm]{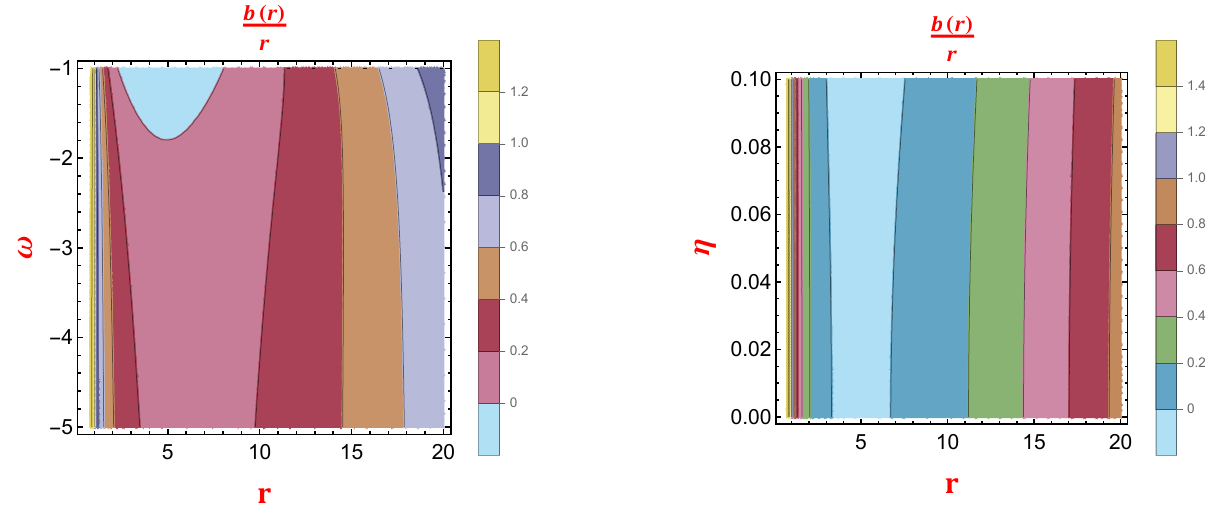}
    \caption{The contour plot illustrates the barotropic EoS, depicting changes in the asymptotic flatness condition for $\omega$ \textit{(depicted on the left)} and $\eta$ \textit{(depicted on the right)} concerning the radial coordinate `$r$' with the redshift function $\phi(r)=\log\left(1+\frac{r_0}{r}\right)$. Additionally, the plot maintains constant values for other parameters, such as $\alpha=-2$, $\beta=0.02$, and $r_0 = 1$.}
    \label{fig17}
\end{figure*}

The function $\phi(r)=\log\left(1+\frac{r_0}{r}\right)$ \cite{N. Godani} serves as another viable option as a redshift function, effectively avoiding the event horizon beyond the wormhole throat. Upon solving equations \eqref{16} and \eqref{17} while incorporating the relation \eqref{19}, we deduce the following first-order differential equation:
\begin{multline}
    b'(r) -\left(\frac{\alpha  (r-r_0) }{r \omega  (r_0+r)}\right)b(r)= \frac{1}{2 \alpha \omega  (r_0+r)}\left(r_0 \left(4 \alpha 
   \right. \right. \\ \left.\left.
    \hspace{0.7cm}  +16 \pi  \eta ^2 (\omega +1)+r^2 (\beta -\beta  \omega )\right)+r^3 (\beta -\beta  \omega )
    \right. \\ \left.
   +16 \pi  \eta ^2  r (\omega +1)\right)
\end{multline}
By integrating the aforementioned equation with the condition $b(r_0) = r_0$, we obtain an explicit expression for the shape function as
\begin{multline}\label{33}
    b(r)= \frac{2^{-\frac{\omega +2}{\omega }} r_0^{-1/\omega } r^{-1/\omega }}{\alpha  \omega  (6 \omega -5)+\alpha} \left(4^{1/\omega } r_0^{1/\omega } r^{\frac{1}{\omega }+1} (\beta  (\omega -1) 
    \right. \\ \left.
  \hspace{0.6cm} \times (r_0+r) (2 r_0 \omega +r_0-2 r \omega +r)+2 \alpha  (\omega  (6 \omega -5)+1))
   \right. \\ \left.
  \hspace{0.5cm} -4 r_0^3 \beta  (\omega -1) (r_0+r)^{2/\omega }\right) + \frac{1}{2 r_0 \alpha  (\omega +1) (\omega  (6 \omega -5)+1)} \\
   \hspace{0.6cm} \times \left( \left((\omega +1) \left(r_0^2 \beta  (\omega -1) (2 \omega +1)-16 \pi  \eta ^2 (\omega  (6 \omega -5)
   \right.\right.\right. \\ \left.\left.\left.
  \hspace{0.7cm} +1)\right)+2 \alpha  (\omega -1) (2 \omega -1) (3 \omega -1)\right) \left(-r (r_0+r) \,
  \right.\right. \\ \left.\left.
  \hspace{0.7cm} \times _2F_1\left(1,2-\frac{1}{\omega };2+\frac{1}{\omega };-\frac{r}{r_0}\right)+r_0^{2-\frac{1}{\omega }} (r_0+r)^{2/\omega } 
 \right.\right. \\ \left.\left.
  \hspace{0.7cm} \times r^{-1/\omega }\, _2F_1\left(1+\frac{1}{\omega },\frac{2}{\omega };2+\frac{1}{\omega };-1\right) \right) \right) \,,
\end{multline}
where $_2F_1$ is the hypergeometric function and it can be defined by $_2F_1(a,b,c,z)=\mathlarger{\sum}\limits_{k=0}^{\infty } \frac{a_k b_k z^k}{k! c_k}$.\\
It is evident that the above expression is not asymptotically flat, which means that the ratio $\frac{b(r)}{r}$ does not tend to $0$ as $r \to \infty$ and graphically presented in the Contour plot \ref{fig17}. As a result, it is challenging to obtain asymptotically flat wormhole solutions for barotropic EoS under a particular logarithmic redshift function for the $f(Q)$ model given in Eq. \eqref{15}.

\subsubsection{$\phi(r)=\frac{1}{r}$}\label{subsubsec3}
\begin{figure*}[t]
    \centering
    \includegraphics[width=17.5cm,height=5cm]{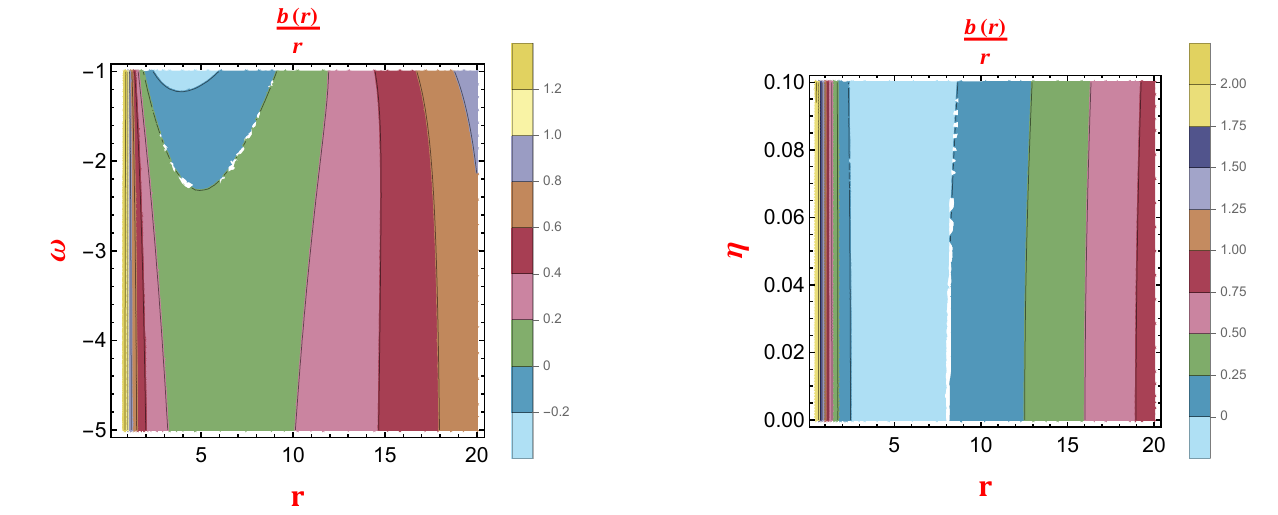}
    \caption{The contour plot illustrates the barotropic EoS, depicting changes in the asymptotic flatness condition for $\omega$ \textit{(depicted on the left)} and $\eta$ \textit{(depicted on the right)} concerning the radial coordinate `$r$' with the redshift function $\phi(r)=\frac{1}{r}$. Additionally, the plot maintains constant values for other parameters, such as $\alpha=-2$, $\beta=0.02$, and $r_0 = 1$.}
    \label{fig18}
\end{figure*}

The function $\phi(r) = \frac{1}{r}$ \cite{S. Kar, F. Parsaei} presents another suitable alternative as a redshift function, given its consistent behavior for $r > 0$, effectively circumventing the event horizon beyond the wormhole throat. Now, by again solving equations \eqref{16} and \eqref{17} while incorporating the relation \eqref{19} for the particular redshift function $\phi(r) = \frac{1}{r}$, the first-order differential equation is given by :
\begin{multline}
    b'(r)-\left(\frac{(r-2)}{r^2 \omega}\right) b(r)=\frac{1}{2 \alpha  r \omega}\left(4 \alpha +r^3 (\beta -\beta  \omega )
    \right.\\ \left.
    +16 \pi  \eta ^2 r (\omega +1)\right)
\end{multline}
By integrating the above first order differentiable equation with the throat condition, we get the shape function as below
\begin{multline}
    b(r)=\frac{2^{-1/\omega } e^{\frac{2}{r \omega }} \left(\frac{1}{r \omega }\right)^{-1/\omega }}{\alpha  \omega ^4} \left( 16 \pi  \eta ^2 \omega ^2 (\omega +1)
     \right. \\ \left.
  \hspace{0.7cm} \times \Gamma \left(\frac{1}{\omega }-1,\frac{2}{r \omega }\right)-4 \beta  (\omega -1) \Gamma \left(\frac{1}{\omega }-3,\frac{2}{r \omega }\right) 
   \right. \\ \left.
  \hspace{0.7cm} +r_0^{-1/\omega } e^{-\frac{2}{r_0 \omega }} \left(\frac{1}{r_0 \omega }\right)^{-1/\omega } r^{1/\omega } \left(\frac{1}{r \omega }\right)^{1/\omega }\left(4 \beta 
  \right.\right. \\ \left.\left.
  \hspace{0.7cm} \times (\omega -1) e^{\frac{2}{r_0 \omega }} \Gamma \left(\frac{1}{\omega }-3,\frac{2}{r_0 \omega }\right)+\frac{1}{\omega -1}\left(\omega ^3 \left(\alpha  
\right.\right.\right.\right. \\ \left.\left.\left.\left.
  \hspace{0.7cm} \times (\omega -1)-8 \pi  \eta ^2 (\omega +1)\right) \left(r_0 2^{1/\omega } \omega  \left(\frac{1}{r_0 \omega }\right)^{1/\omega }
\right.\right.\right.\right. \\ \left.\left.\left.\left.
  \hspace{0.7cm} -2 e^{\frac{2}{r_0 \omega }} \Gamma \left(\frac{1}{\omega },\frac{2}{r_0 \omega }\right)\right)\right)\right)+2 \alpha  \omega ^3 \Gamma \left(\frac{1}{\omega },\frac{2}{r \omega }\right)\right) 
\end{multline}
Further, visual representation of the above shape function is shown in the Contour plot \ref{fig18} and it is clear that the shape function obtained for the redshift function $\phi(r)=\frac{1}{r}$ case is not asymptotically flat.

\subsection{Anisotropic EoS}\label{subsec2}
In this specific segment, we focus on exploring the anisotropic energy-momentum tensor fluid relation within the context of asymptotically flat wormhole solutions. Unlike scenarios where $p_r$ equals $p_t$, we delve into cases where a disparity exists between these radial and tangential pressures. To elucidate this relationship between $p_r$ and $p_t$, we refer to the formulation presented in prior works \cite{F. Rahaman 1, P. H. R. S. Moraes} as
\begin{equation}\label{25}
p_t = n\,p_r\,.
\end{equation}
In this context, the parameter $n$ assumes significance, as it governs the characteristics of the anisotropic fluid under examination. It is imperative to note that in our study, $n$ cannot hold the value of one, denoted as $n \neq 1$. This restriction is crucial because if $n$ were equal to one, it would simplify into that of a perfect fluid. Moreover, we study the case where $n=1$ in the next subsection \ref{subsec3}. Hence, our analysis exclusively pertains to scenarios where $n$ deviates from unity, thereby retaining the complexities associated with anisotropic fluid behavior by employing redshift functions, as discussed in Subsubsections \ref{subsubsec4}-\ref{subsubsec6}.\\ 
\subsubsection{$\phi(r)=c$}\label{subsubsec4}
\begin{figure*}[t]
    \centering
    \includegraphics[width=14.5cm,height=5cm]{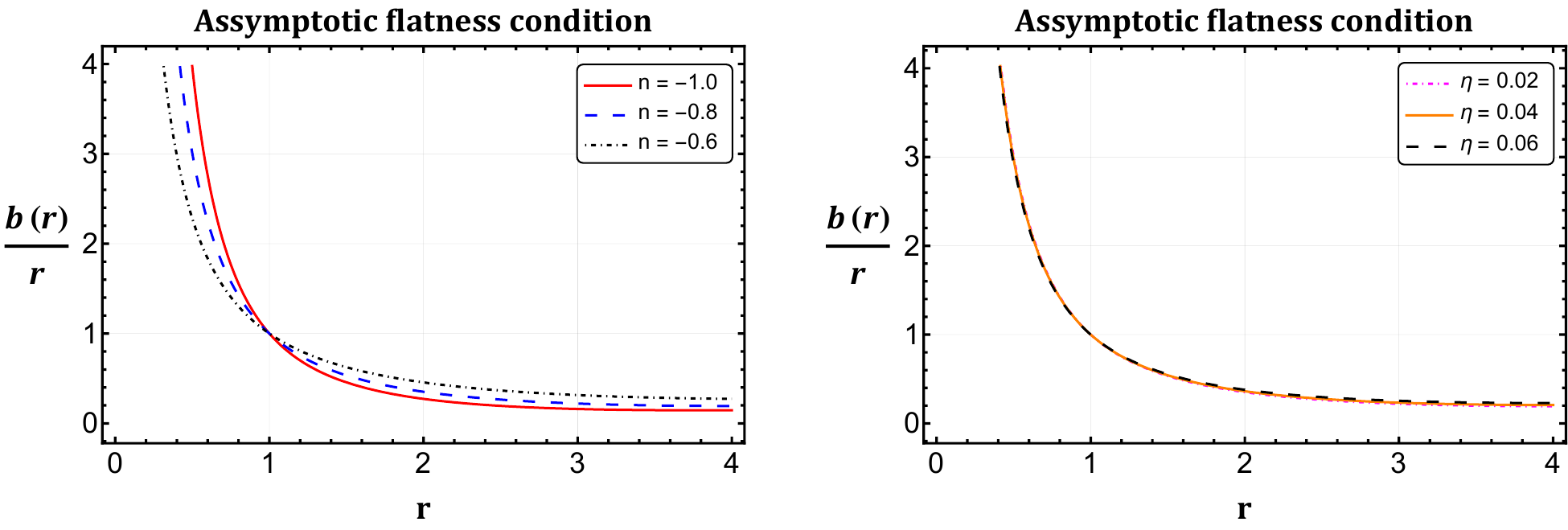}
    \caption{The plot illustrates the anisotropic EoS, depicting changes in the asymptotic flatness condition for various values of $n$ \textit{(depicted on the left)} and $\eta$ \textit{(depicted on the right)} concerning the radial coordinate `$r$'. Additionally, the plot maintains constant values for other parameters, such as $\alpha=-2$, $\beta=0.02$, and $r_0 = 1$.}
    \label{fig7}
\end{figure*}

\begin{figure*}[t]
    \centering
    \includegraphics[width=14.5cm,height=5cm]{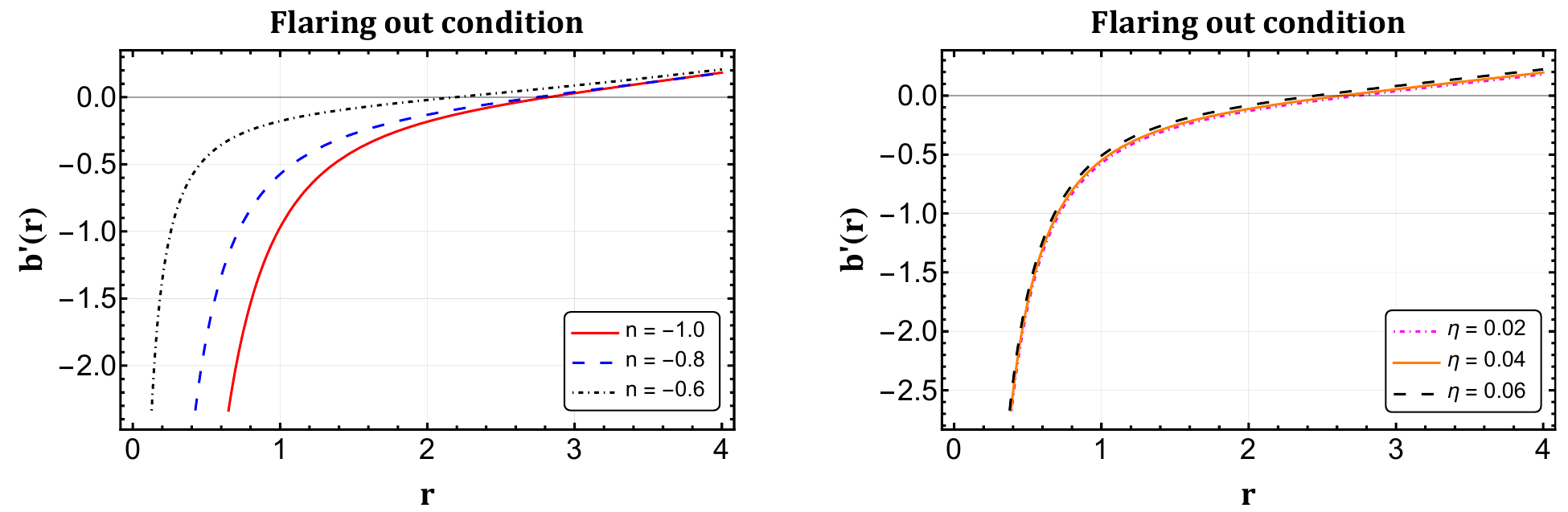}
    \caption{The plot illustrates the anisotropic EoS, depicting changes in the flaring-out condition for various values of $n$ \textit{(depicted on the left)} and $\eta$ \textit{(depicted on the right)} concerning the radial coordinate `$r$'. Additionally, the plot maintains constant values for other parameters, such as $\alpha=-2$, $\beta=0.02$, and $r_0 = 1$.}
    \label{fig8}
\end{figure*}

\begin{figure*}[t]
\centering
\includegraphics[width=14.5cm,height=5cm]{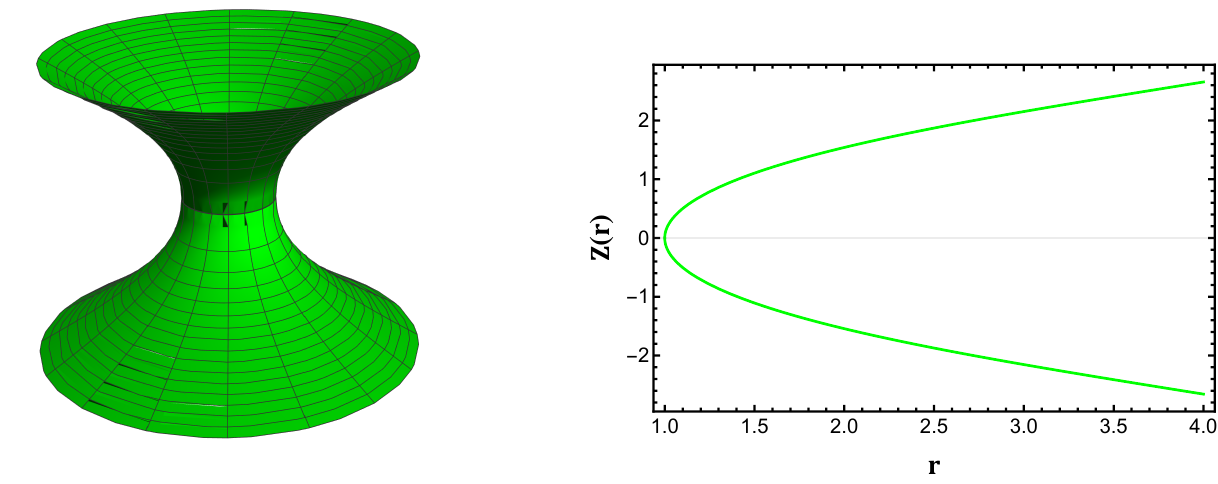}
\caption{The figure displays the embedding diagram for the anisotropic EoS. Additionally, the figure maintains constant values for other parameters, such as $\alpha=-2$, $\beta=0.02$, $\eta=0.02$, $n=-0.8$ and $r_0 = 1$.}
\label{fig14}
\end{figure*}

\begin{figure*}[t]
    \centering
    \includegraphics[width=14.5cm,height=5cm]{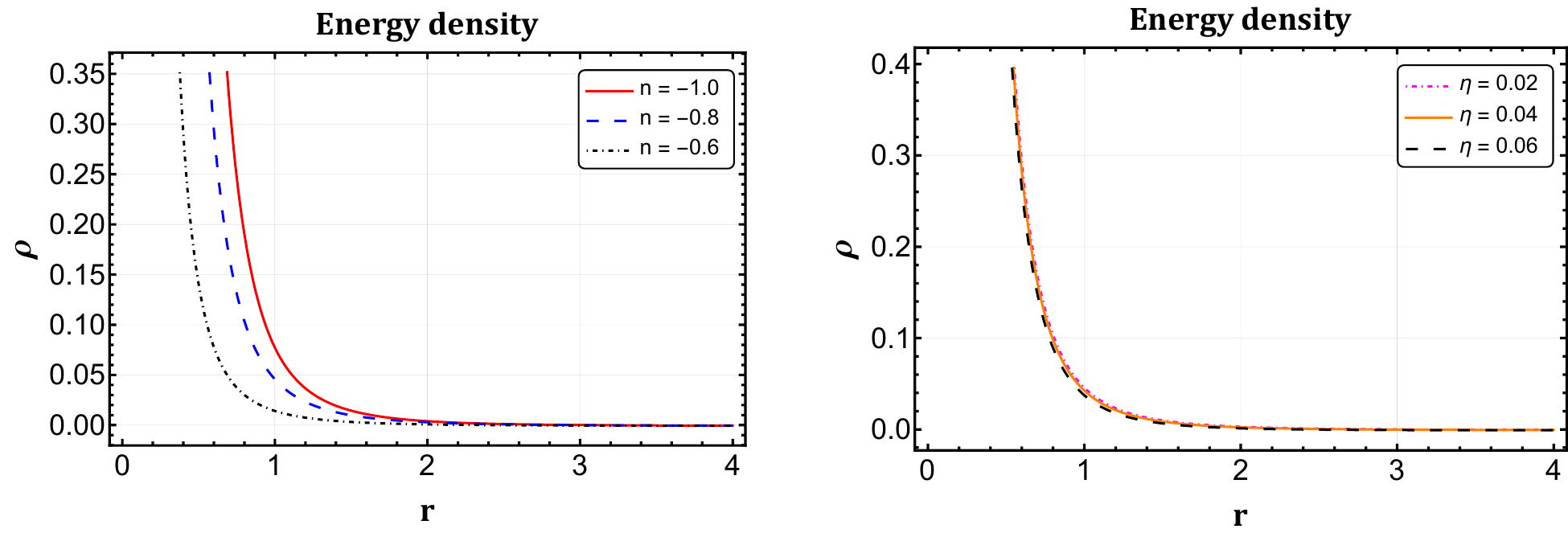}
    \caption{The plot illustrates the anisotropic EoS, depicting changes in the energy density for various values of $n$ \textit{(depicted on the left)} and $\eta$ \textit{(depicted on the right)} concerning the radial coordinate `$r$'. Additionally, the plot maintains constant values for other parameters, such as $\alpha=-2$, $\beta=0.02$, and $r_0 = 1$.}
    \label{fig9}
\end{figure*}

\begin{figure*}[t]
    \centering
    \includegraphics[width=14.5cm,height=5cm]{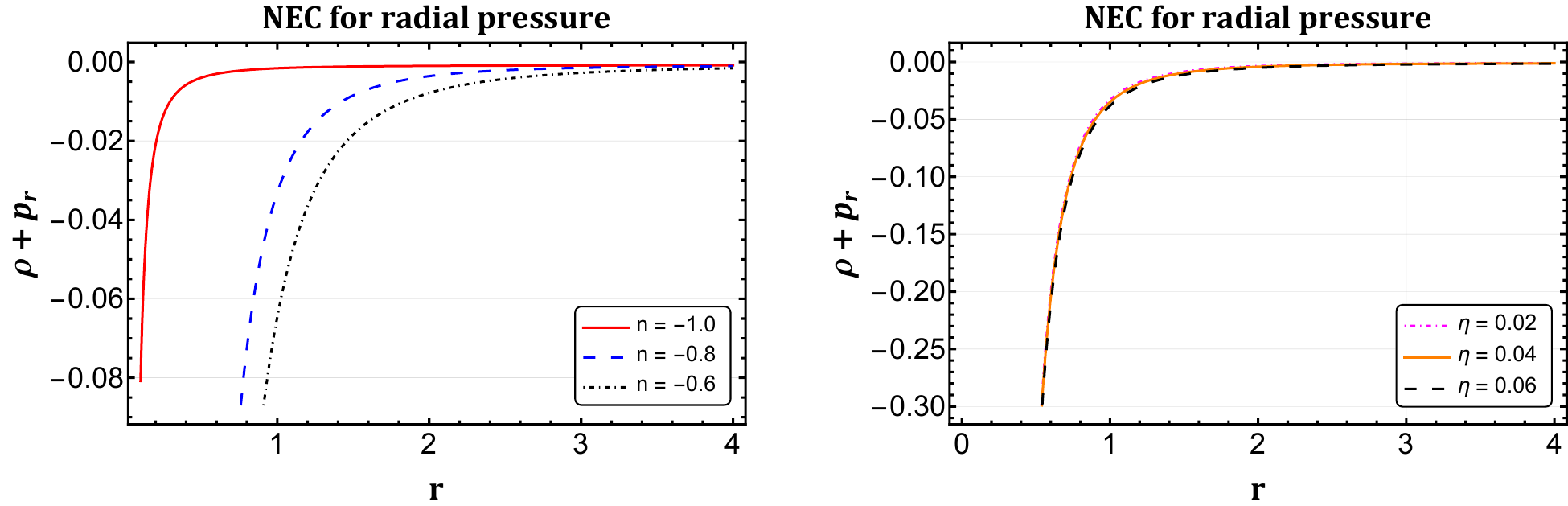}
    \caption{The plot illustrates the anisotropic EoS, depicting changes in the NEC for radial pressure for various values of $n$ \textit{(depicted on the left)} and $\eta$ \textit{(depicted on the right)} concerning the radial coordinate `$r$'. Additionally, the plot maintains constant values for other parameters, such as $\alpha=-2$, $\beta=0.02$, and $r_0 = 1$.}
    \label{fig10}
\end{figure*}

\begin{figure*}[t]
    \centering
    \includegraphics[width=14.5cm,height=5cm]{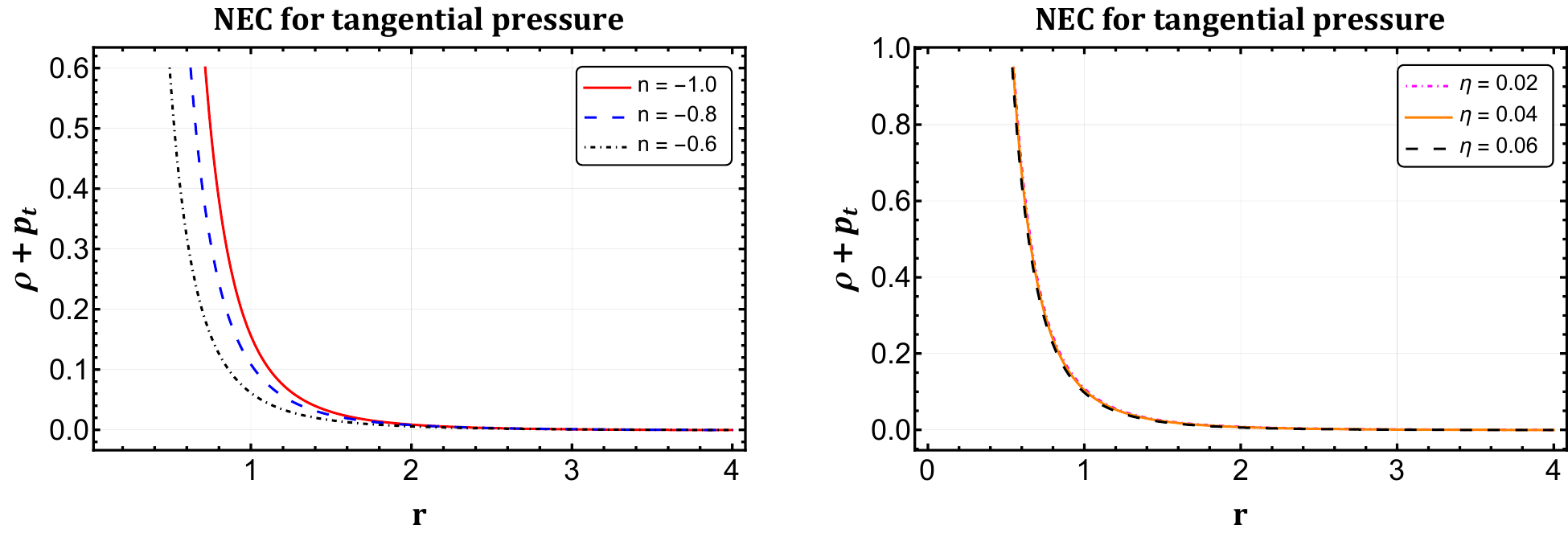}
    \caption{The plot illustrates the anisotropic EoS, depicting changes in the NEC for tangential pressure for various values of $n$ \textit{(depicted on the left)} and $\eta$ \textit{(depicted on the right)} concerning the radial coordinate `$r$'. Additionally, the plot maintains constant values for other parameters, such as $\alpha=-2$, $\beta=0.02$, and $r_0 = 1$.}
    \label{fig11}
\end{figure*}

\begin{figure*}[t]
    \centering
    \includegraphics[width=14.5cm,height=5cm]{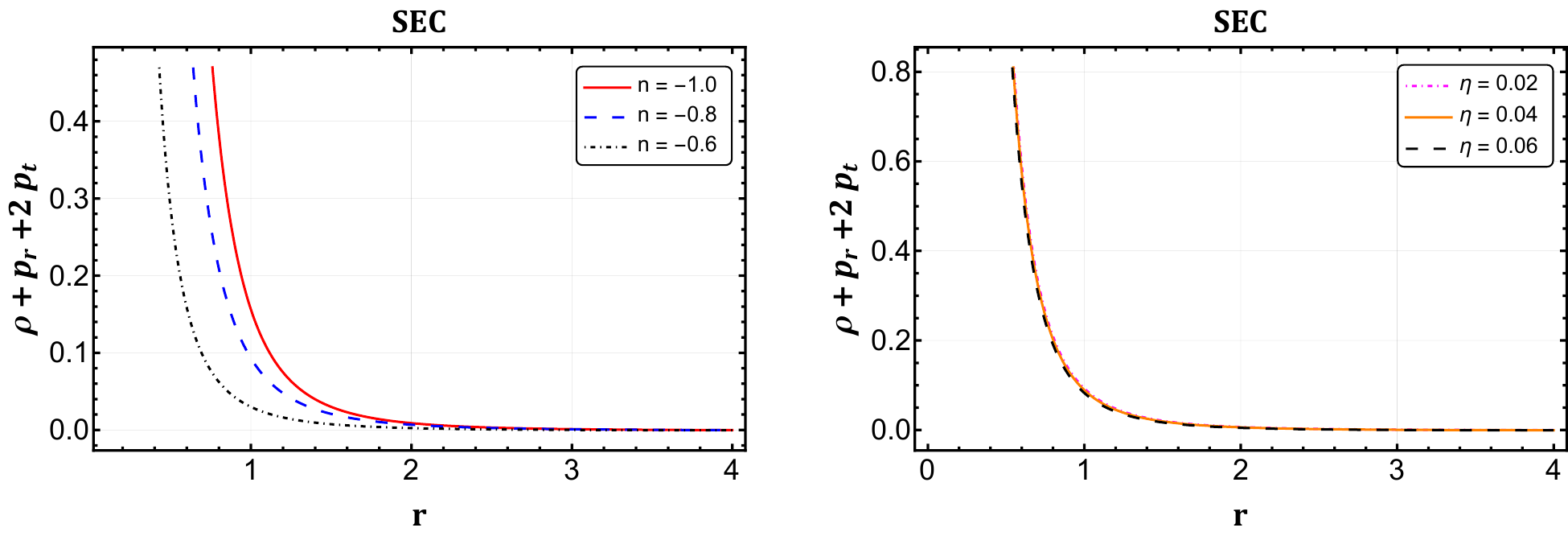}
    \caption{The plot illustrates the anisotropic EoS, depicting changes in the SEC for various values of $n$ \textit{(depicted on the left)} and $\eta$ \textit{(depicted on the right)} concerning the radial coordinate `$r$'. Additionally, the plot maintains constant values for other parameters, such as $\alpha=-2$, $\beta=0.02$, and $r_0 = 1$.}
    \label{fig12}
\end{figure*}

In this specific subsubsection, we use Eqs. \eqref{17} and \eqref{18} with consideration of the relationship \eqref{25}, we obtain the subsequent first-order differential equation:
\begin{multline}\label{26}
b'(r) -\left(\frac{2n}{r}+\frac{1}{r}\right) b(r)= \frac{\beta  n r^3+16 \pi  \eta ^2 n r-\beta  r^3}{\alpha  r}    \,.
\end{multline}
Upon integrating the aforementioned equation with the throat condition $b(r_0) = r_0$, we derive the explicit form of the shape function as:
\begin{multline}\label{27}
b(r) = \frac{-r}{2 \alpha }\left(-r_0^{-2 n} r^{2 n}\left(r_0^2 \beta +2 \alpha +16 \pi  \eta ^2\right)+16 \pi  \eta ^2
\right. \\ \left.
+\beta  r^2\right)\,.
\end{multline}
Now, we transition into the visual analysis of the shape function, where we explore the pivotal conditions governing the existence of a wormhole. To facilitate this examination, we meticulously select appropriate parameters. We begin by presenting Fig. \ref{fig7}, showcasing the asymptotic behavior of the shape function with respect to the various values of parameter $\omega$ (depicted in the left plot) and $\eta$ (displayed in the right plot). This graphical representation confirms the trend where, as the radial distance increases, the ratio $\frac{b(r)}{r}$ tends towards $0$, thus showing the satisfaction of the asymptotic flatness condition. Furthermore, we ensure the fulfillment of the crucial flaring-out condition $b'(r_0) < 1$ at the wormhole throat $r = r_0$, as demonstrated in Fig. \ref{fig8}, utilizing the various values of the same parameters $\omega$ and $\eta$. Additionally, we provide a comprehensive visualization of the embedding diagram in Fig. \ref{fig14}.\\
By utilizing Eq. \eqref{27} within Eqs. \eqref{16}-\eqref{18}, one can derive the expressions for the energy density, radial pressure, and tangential pressure as follows:
\begin{multline}\label{28}
\rho= \frac{1}{16 \pi  r^2} \left((2 n+1) r_0^{-2 n} r^{2 n} \left(r_0^2 \beta +2 \alpha +16 \pi  \eta ^2\right)
\right. \\ \left.
-32 \pi  \eta ^2-2 \beta  r^2\right)\,,
\end{multline}
\begin{multline}\label{29}
    p_r= \frac{1}{16 \pi}\left(r_0^{-2 n} r^{2 n-2} \left(r_0^2 \beta +2 \alpha +16 \pi  \eta ^2\right)\right)\,,
\end{multline}
\begin{multline}\label{30}
    p_t = \frac{1}{16 \pi}\left(n r_0^{-2 n} r^{2 n-2} \left(r_0^2 \beta +2 \alpha +16 \pi  \eta ^2\right)\right)\,.
\end{multline}
Moving forward, we showcase the energy density's behavior using Fig. \ref{fig9}, depicting its gradual decreasing behavior throughout Space-time concerning the parameters $n$ and $\eta$ across the radial coordinate $r$. Moreover, it is imperative to account for the NEC concerning the radial and tangential pressures at the wormhole throat, which can be expressed as: 
\begin{equation}
\left(\rho + p_r\right)_{\text{at}\,\, r=r_0}=\frac{r_0^2 \beta  n+2 \alpha +2 \alpha  n+16 \pi  \eta ^2 n}{8 \pi  r_0^2}\,,
\end{equation}
\begin{multline}
\hspace{-0.5cm}\left(\rho + p_t\right)_{\text{at}\,\, r=r_0}=\frac{(3 n-1) \left(r_0^2 \beta +16 \pi  \eta ^2\right)+\alpha  (6 n+2)}{16 \pi  r_0^2}\,.
\end{multline}
Further, Figs. \ref{fig10} and \ref{fig11} visually elucidate the NEC's radial and tangential pressure aspects at the throat. Fig. \ref{fig10} reveals the NEC's negative increase trend in radial pressure with increasing radial distance, while Fig. \ref{fig11} highlights the NEC's positive trend in tangential pressure. The violation of NEC in this barotropic scenario suggests the likely presence of exotic matter required for throat stability. Extending our analysis to the SEC, Fig. \ref{fig12} demonstrates its satisfaction near the throat for both parameters, providing a comprehensive visualization of the SEC's behavior. 
\begin{table*}[t]
\begin{tabular}{ccccccc}
\\ \hline \hline

 & \multicolumn{2}{p{5cm}}{Barotropic EoS ($p_r=\omega \rho$)}                                   & \multicolumn{2}{p{5cm}}{Anisotropic EoS ($p_t=n p_r$)}                       & \multicolumn{1}{l}{Isotropic EoS ($p_t=p_r$)} 
 \\ \hline\hline
 & \multicolumn{1}{c}{$\omega$} & \multicolumn{1}{c}{ $\eta$} & \multicolumn{1}{c}{$n$} & \multicolumn{1}{c}{$\eta$} & \multicolumn{1}{c}{-}         \\ \hline
$\rho$   & Validated    & Validated      &  Validated  &   Validated    &       -       \\
$\rho+p_r$  & Failed    & Failed      &  Failed  &   Failed    &     -       \\
$\rho+p_t$  & Validated    & Validated      &  Validated  &   Validated    &        -       \\
$\rho-p_r$  & Validated    & Validated      &  Validated  &   Validated    &    -           \\
$\rho-p_t$  & Failed    & Failed     &  Failed  &   Failed    &   -            \\
$\rho+p_r+2p_t$  & Validated    & Validated      &  Validated  &   Validated    &      - 
\\ \hline \hline
\end{tabular}
\caption{Summary of energy conditions for results of linear EoS cases under constant redshift function}
\label{table:1}
\end{table*}

\subsubsection{$\phi(r)=\log\left(1+\frac{r_0}{r}\right)$}\label{subsubsec5}
\begin{figure*}[t]
    \centering
    \includegraphics[width=17.5cm,height=5cm]{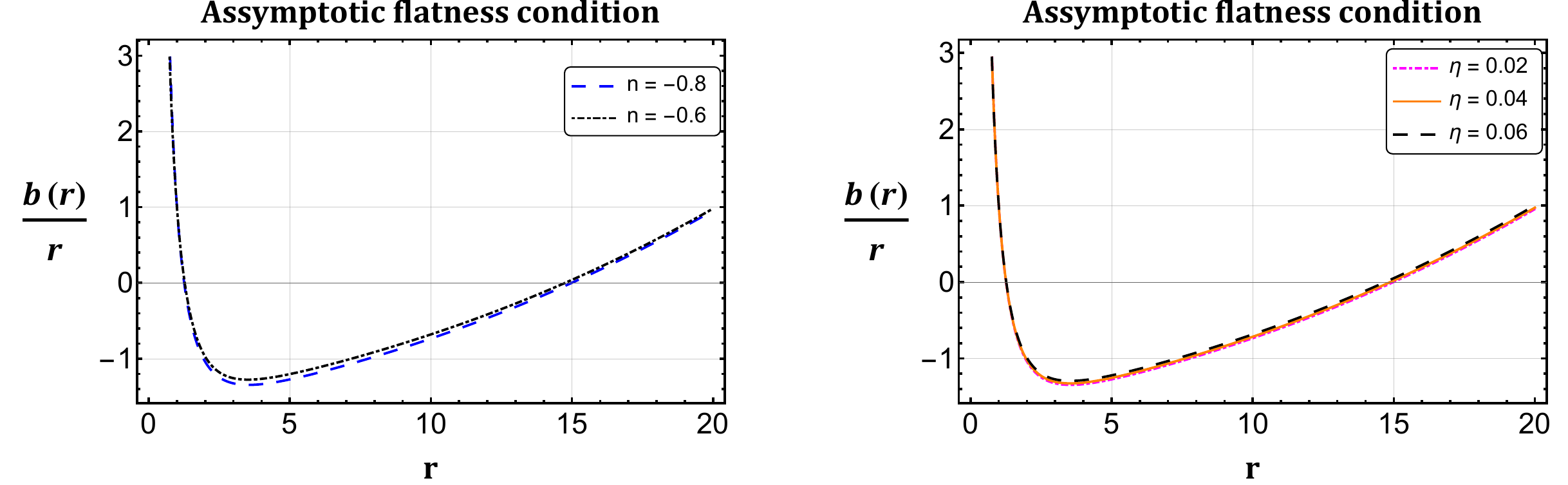}
    \caption{The contour plot illustrates the anisotropic EoS, depicting changes in the asymptotic flatness condition for $n$ \textit{(depicted on the left)} and $\eta$ \textit{(depicted on the right)} concerning the radial coordinate `$r$' with the redshift function $\phi(r)=\log\left(1+\frac{r_0}{r}\right)$. Additionally, the plot maintains constant values for other parameters, such as $\alpha=-2$, $\beta=0.02$, and $r_0 = 1$.}
    \label{fig23}
\end{figure*}

Utilizing this specific form of the redshift function $\phi(r)=\log\left(1+\frac{r_0}{r}\right)$ and the Eqs. \eqref{17} and \eqref{18} with the relation \eqref{25}, we obtain the subsequent first-order differential equation as
\begin{multline}
    b'(r)-\left(\frac{\alpha (-2 r_0 (n+1)+2 n r+r)}{\alpha  r^2} \right) b(r)= \frac{1}{\alpha  r}\left(r_0 \left(\alpha 
    \right.\right. \\ \left.\left.
    \hspace{1cm}\times (4 n+2)+16 \pi  \eta ^2 n+\beta  (n-1) r^2\right)+\beta  (n-1) r^3
    \right. \\ \left.
    +16 \pi  \eta ^2 n r\right)
\end{multline}
Further, we get the explicit form of the shape function by integrating the above equation as
\begin{multline}
    b(r)=\frac{r^{2 n+1} e^{\frac{2 r_0 (n+1)}{r}}}{\alpha  (1-2 n) (n+1)}\left(\frac{1}{2}r_0^{-2 n}e^{-2 (n+1)}\left(r_0^2 \beta +n 
   \right.\right. \\ \left.\left. 
   \hspace{0.6cm} \times \left(-\left(r_0^2 \beta  (6 n+5)\right)+\alpha  (4 n-2)+16 \pi  \eta ^2 (2 n-1)\right)
   \right.\right. \\ \left.\left. 
   \hspace{0.6cm} +4 e^{2 n+2} n E_{1-2 n}(2 (n+1)) \left(2 r_0^2 \beta  (n+1)^2+  \left(4 n^2-1\right)
    \right.\right.\right. \\ \left.\left.\left. 
   \hspace{0.6cm}+8 \pi  \eta ^2 \left(4 n^2-1\right)\right)\right)+r^{-2 n}\left(\frac{1}{2} e^{-\frac{2 r_0 (n+1)}{r}} \left(\beta  (n+1) r 
    \right.\right.\right. \\ \left.\left.\left. 
   \hspace{0.6cm} \times (4 r_0 n+2 n r-r)+\alpha  \left(2-8 n^2\right)+16 \pi  \eta ^2 n (1-2 n)\right)
  \right.\right. \\ \left.\left. 
   \hspace{0.6cm}  -2 n E_{1-2 n}\left(\frac{2 r_0 (n+1)}{r}\right) \left(2 r_0^2 \beta  (n+1)^2+\alpha  \left(4 n^2-1\right)
   \right.\right.\right. \\ \left.\left.\left. 
   \hspace{0.6cm}  +8 \pi  \eta ^2 \left(4 n^2-1\right)\right)\right)\right)+\frac{1}{2 r_0 \alpha  (n+1) (n (6 n-5)+1)}\\
   \hspace{0.7cm} \times \left((n+1) \left(r_0^2 \beta  (n-1) (2 n+1)-16 \pi  \eta ^2 (n (6 n-5)
 \right.\right. \\ \left.\left. 
   \hspace{0.6cm} +1)\right)+2 \alpha  (n-1) (2 n-1) (3 n-1)\right)\left(r_0^{2-\frac{1}{n}} (r_0+r)^{2/n}
  \right. \\ \left.
  \hspace{0.6cm} \times r^{-1/n} \, _2F_1\left(1+\frac{1}{n},\frac{2}{n};2+\frac{1}{n};-1\right)-r (r_0+r)
\right. \\ \left.
  \hspace{0.6cm} \times _2F_1\left(1,2-\frac{1}{n};2+\frac{1}{n};-\frac{r}{r_0}\right)\right)\,,
\end{multline}
where, ``E" is an exponential integral function and defined by $E_n(z)=\int _1^{\infty }\frac{e^{-zt}}{t^n}dt $.\\
Moreover, a visual representation of the above shape function is shown in Fig. \ref{fig23}, and clearly, the asymptotic flatness condition is not satisfied here for this redshift choice. Also, from the above expression, it is clear that the shape function is not defined for $n=-1, \frac{1}{2}$. Hence, the plot for $n=-1$ is not visible in the Fig. \ref{fig23}.

\subsubsection{$\phi(r)=\frac{1}{r}$}\label{subsubsec6}
By employing this particular form of the redshift function $\phi(r)=\frac{1}{r}$ alongside Eqs. \eqref{17} and \eqref{18}, incorporating the relation \eqref{25}, we derive the following subsequent first-order differential equation:
\begin{multline}
    b'(r)-\left(\frac{r (2 n (r-2)+r-3)-2}{(r-1) r^2}\right) b(r)= \frac{1}{\alpha  (r-1) r^2}\\
    \times \left(\beta  (n-1) r^5+16 \pi  \eta ^2 n r^3+2 \alpha  r (2 n r+r+1)\right)
\end{multline}
Furthermore, the above-mentioned differential equation does not have a solution in closed form. Even after extensive attempts using computational tools like \textit{Mathematica}, no closed-form solution has been attained, indicating that it cannot be expressed in terms of elementary functions. Consequently, we are unable to make any conclusive remarks regarding the condition of asymptotic flatness.

\subsection{Isotropic EoS}\label{subsec3}
Typically, the energy-momentum tensor of asymptotically flat wormholes in GR exhibits anisotropy, where the radial pressure $p_r$ differs from the tangential pressure $p_t$. We have previously delved into wormhole solutions for this anisotropic EoS (\ref{subsec2}) scenario. In this section, we shift our focus to investigating wormhole solutions under the following isotropic relation:
\begin{equation}\label{31}
    p_r=p_t\,.
\end{equation}
Now, by utilizing Eqs. \eqref{17} and \eqref{18} while incorporating the relationship \eqref{31}, we arrive at the following first-order differential equation:
\begin{equation}\label{32}
b'(r)-\left(\frac{3}{r}\right) b(r)= \frac{16 \pi  \eta ^2 r}{\alpha  r}\,.
\end{equation}
By integrating the above-mentioned equation with the condition $b(r_0) = r_0$, we derive an explicit expression of the shape function as
\begin{equation}\label{33}
b(r)= \frac{r}{\alpha} \left(\frac{r^2 \left(\alpha +8 \pi  \eta ^2\right)}{r_0^2}-8 \pi  \eta ^2\right)\,.
\end{equation}
It is evident that the above expression is not asymptotically flat, which means that the ratio $\frac{b(r)}{r}$ does not tend to $0$ as $r \to \infty$. As a result, it is challenging to obtain asymptotically flat wormhole solutions for isotropic pressure under a constant redshift function for the $f(Q)$ model given in Eq. \eqref{15}. This type of result can be found in the reference \cite{R. Solanki}. \\
Moreover, our findings are consistent with those discussed in Subsubsections (\ref{subsubsec2}-\ref{subsubsec3}) and (\ref{subsubsec5}-\ref{subsubsec6}) when examining the redshift functions $\phi(r)=\log\left(1+\frac{r_0}{r}\right)$ and $\phi(r)=\frac{1}{r}$. Specifically, we determined that the shape functions derived from these redshift functions do not meet the criteria for asymptotic flatness.

\section{Equilibrium Conditions}
\label{sec5}
This section utilizes the generalized Tolman-Oppenheimer-Volkov (TOV) equation \cite{Oppenheimer, Gorini, Kuhfittig} to assess the stability of our wormhole solutions. The generalized TOV equation can be expressed as
\begin{eqnarray}\label{51}
\frac{\varpi^{'}}{2}(\rho+p_r)+\frac{dp_r}{dr}+\frac{2}{r}(p_r-p_t)=0,
\end{eqnarray}
where $\varpi=2\phi(r)$.\\
Depending on the anisotropic matter distribution, the hydrostatic, gravitational, and anisotropic forces are outlined as follows:
\begin{equation}\label{52}
F_H=-\frac{dp_r}{dr}, ~~F_G=-\frac{\varpi^{'}}{2}(\rho+p_r), ~F_A=\frac{2}{r}(p_t-p_r).    
\end{equation}
To achieve equilibrium in the wormhole solutions, it is essential for the equation $F_H+F_G+F_A=0$ to hold. In this study, we have assumed a constant redshift function $\phi(r)$, which implies that the gravitational contribution $F_G$ vanishes in the equilibrium equation. Thus, the equilibrium equation simplifies to:
\begin{equation}
\label{52a}
F_H+F_A=0.
\end{equation}
Utilizing Eqs. \eqref{22}, \eqref{23}, and \eqref{52}, we obtain the following equations for the hydrostatic and anisotropic forces for the linear model \eqref{15} using barotropic EoS case, i.e., $p_r=\omega \rho$ as
\begin{multline}
F_H=\frac{3}{16 \pi  r^4} \left(2\alpha \left(\frac{8 \pi  \eta ^2 r (\omega +1)}{\alpha  (\omega -1)}-\frac{\beta  r^3 (\omega -1)}{2 \alpha  (3 \omega -1)}
\right.\right. \\ \left.\left.
+r_0^{-1/\omega } r^{1/\omega } \left(\frac{r_0^3 \beta  (\omega -1)}{2 \alpha  (3 \omega -1)}-\frac{8 \pi  r_0 \eta ^2 (\omega +1)}{\alpha  (\omega -1)}+r_0\right)\right)
\right. \\ \left.
+\beta  r^3+16 \pi  \eta ^2 r\right)-\frac{1}{16 \pi  r^3}\left(16 \pi  \eta ^2+3 \beta  r^2+2\alpha 
\right. \\ \left.
\times \left( \frac{8 \pi  \eta ^2 (\omega +1)}{\alpha  (\omega -1)}-\frac{3 \beta  r^2 (\omega -1)}{2 \alpha  (3 \omega -1)}+\frac{r_0^{-1/\omega } r^{\frac{1}{\omega }-1}}{\omega}
\right.\right. \\ \left.\left.
\times \left(\frac{r_0^3 \beta  (\omega -1)}{2 \alpha  (3 \omega -1)}-\frac{8 \pi  r_0 \eta ^2 (\omega +1)}{\alpha  (\omega -1)}+r_0\right)\right) \right),
\end{multline}
\begin{multline}
F_A=\frac{2}{r}\left(\frac{-1}{16 \pi  r^3} \left(2\alpha \left(\frac{8 \pi  \eta ^2 r (\omega +1)}{\alpha  (\omega -1)}-\frac{\beta  r^3 (\omega -1)}{2 \alpha  (3 \omega -1)}
\right.\right.\right. \\ \left.\left.\left.
+r_0^{-1/\omega } r^{1/\omega } \left(\frac{r_0^3 \beta  (\omega -1)}{2 \alpha  (3 \omega -1)}-\frac{8 \pi  r_0 \eta ^2 (\omega +1)}{\alpha  (\omega -1)}+r_0\right)\right)
\right.\right. \\ \left.\left.
+\beta  r^3+16 \pi  \eta ^2 r\right) + \frac{1}{16 \pi  r^3}\left(\alpha \left(\frac{-8 \pi  \eta ^2 r (\omega +1)}{\alpha  (\omega -1)}
\right.\right.\right. \\ \left.\left.\left.
+\frac{\beta  r^3 (\omega -1)}{2 \alpha  (3 \omega -1)}-r_0^{-1/\omega } r^{1/\omega } \left(-\frac{8 \pi  r_0 \eta ^2 (\omega +1)}{\alpha  (\omega -1)}+r_0
\right.\right.\right.\right. \\ \left.\left.\left.\left.
+\frac{r_0^3 \beta  (\omega -1)}{2 \alpha  (3 \omega -1)}\right) + r\left(\frac{8 \pi  \eta ^2 (\omega +1)}{\alpha  (\omega -1)}-\frac{3 \beta  r^2 (\omega -1)}{2 \alpha  (3 \omega -1)}
\right.\right.\right.\right. \\ \left.\left.\left.\left.
+\frac{r_0^{-1/\omega } r^{\frac{1}{\omega }-1}}{\omega}\left(\frac{r_0^3 \beta  (\omega -1)}{2 \alpha  (3 \omega -1)}-\frac{8 \pi  r_0 \eta ^2 (\omega +1)}{\alpha  (\omega -1)}+r_0\right)\right)\right)
\right.\right. \\ \left.\left.
+\beta  r^3\right)\right).
\end{multline}
Also, for the anisotropic EoS case, i.e., $p_t=n p_r$, considering Eqs. \eqref{29}, \eqref{30}, and \eqref{52}, the equations for the hydrostatic and anisotropic forces are given as
\begin{equation}
F_H=\frac{-1}{16 \pi }\left((2 n-2) r_0^{-2 n} r^{2 n-3} \left(r_0^2 \beta +2 \alpha +16 \pi  \eta ^2\right)\right),
\end{equation}
\begin{multline}
  F_A=\frac{2}{r}\left(\frac{n r_0^{-2 n} r^{2 n-2}}{16 \pi } \left(r_0^2 \beta +2 \alpha +16 \pi  \eta ^2\right)
  \right. \\ \left.
-\frac{r_0^{-2 n} r^{2 n-2}}{16 \pi } \left(r_0^2 \beta +2 \alpha +16 \pi  \eta ^2\right)\right).  
\end{multline}

\begin{figure*}[t]
    \centering
    \includegraphics[width=14.5cm,height=10cm]{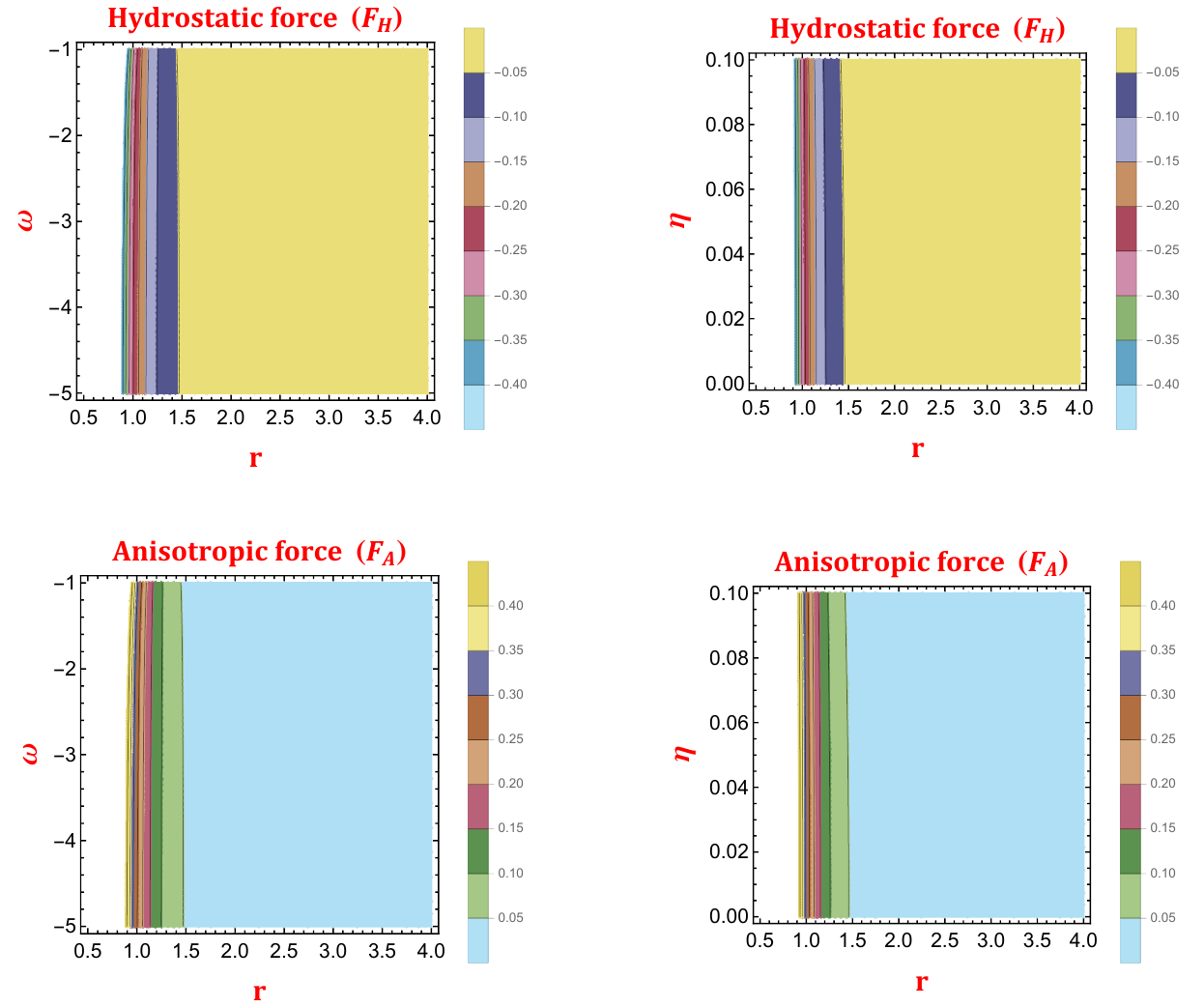}
    \caption{The contour plot illustrates the barotropic EoS ($p_r=\omega \rho$), depicting changes in the hydrostatic and anisotropic forces for $\omega$ \textit{(depicted on the left)} and $\eta$ \textit{(depicted on the right)} concerning the radial coordinate `$r$'. Additionally, the plot maintains constant values for other parameters, such as $\alpha=-2$, $\beta=0.02$, and $r_0 = 1$.}
    \label{fig15}
\end{figure*}
\begin{figure*}[t]
    \centering
    \includegraphics[width=14.5cm,height=12cm]{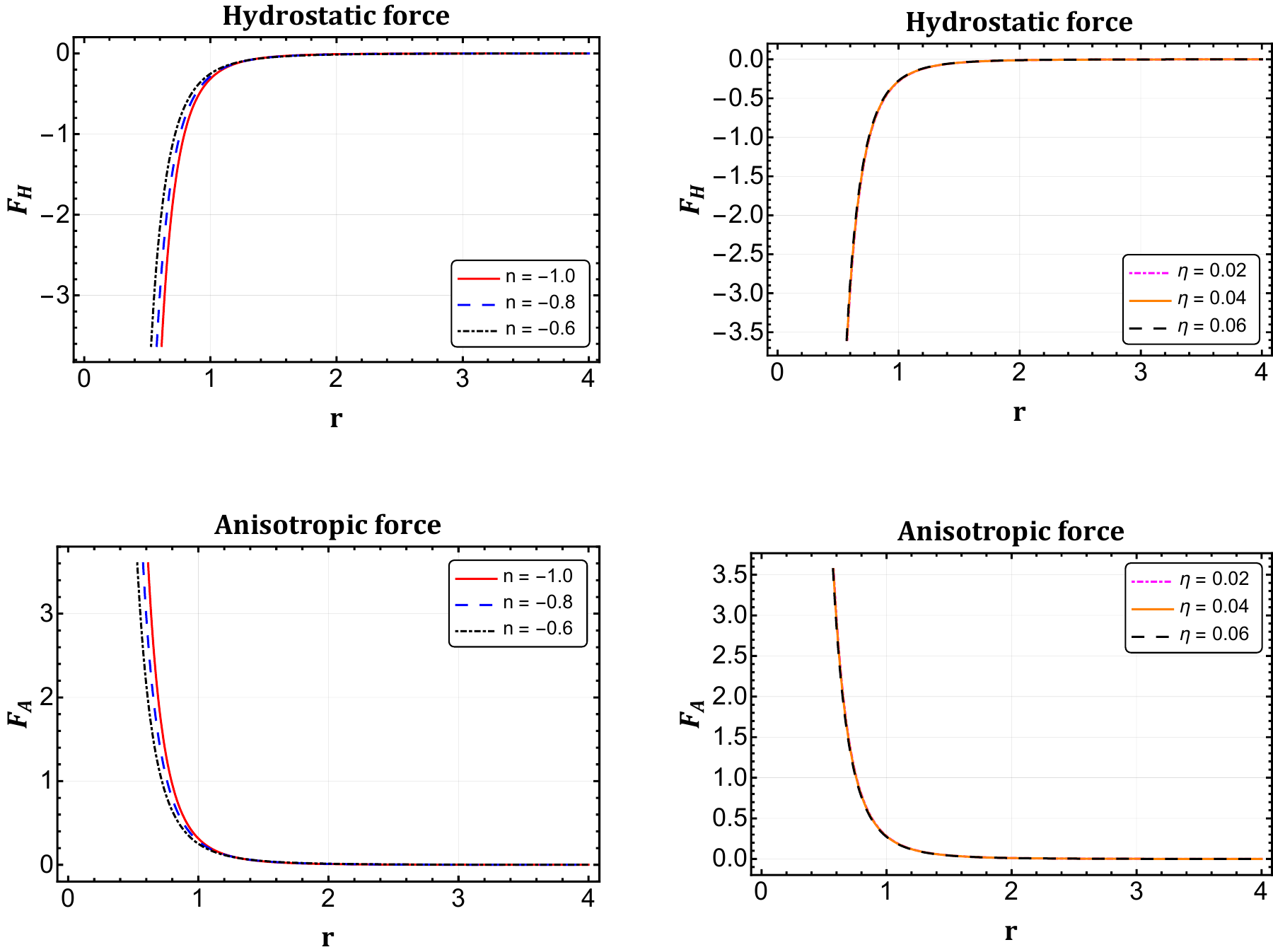}
    \caption{The plot illustrates the anisotropic EoS ($p_t=n p_r$), depicting changes in the hydrostatic and anisotropic forces for various values of $n$ \textit{(depicted on the left)} and $\eta$ \textit{(depicted on the right)} concerning the radial coordinate `$r$'. Additionally, the plot maintains constant values for other parameters, such as $\alpha=-2$, $\beta=0.02$, and $r_0 = 1$.}
    \label{fig16}
\end{figure*}
The hydrostatic and anisotropic forces are visually represented in Figs. \ref{fig15} and \ref{fig16} for both the cases. A notable observation from the graphs is that despite their opposite directions, both forces exhibit similar patterns. This symmetrical balance between the forces suggests the stability of the wormhole solutions obtained in our study.

\section{Conclusion}\label{sec6}
Wormholes serve as hypothetical tunnels connecting two distinct spatial regions separated by a spacelike interval within the same Space-time manifold. Despite being a fascinating theoretical concept, they have yet to be observed. In GR, the wormhole solutions of interest, particularly those that are stable and traversable, encounter a known issue where energy conditions are violated. The necessity for exotic matter, as dictated by standard model extensions within classical GR, poses a significant hurdle. To address this challenge, researchers have turned to modified versions of gravity. Exploring wormholes within modified theories of gravity offers a promising avenue to overcome the issue of energy condition violations. This approach enables the derivation of stable solutions without compromising the fundamental energy conditions. By delving into modified gravity frameworks, researchers aim to refine our understanding of wormholes and potentially pave the way for their realization in the Universe.\\
In our current investigation, we have thoroughly explored wormhole solutions within the framework of $f(Q)$ gravity under the influence of a global monopole charge. This extensive study involved the utilization of three distinct linear EoS juxtaposed against the computed energy density within the $f(Q)$ gravity framework. Our analysis has resulted in a collection of viable and internally consistent solutions to the field equations, specifically tailored to describe a Morris and Thorne wormhole (Eq. \eqref{11}) with linear EoS. These linear EoS comprise the barotropic (Eq. \eqref{19}), anisotropic (Eq. \eqref{25}), and isotropic (Eq. \eqref{31}) cases. The functional forms of the shape functions have been derived, showcasing remarkable accuracy within the specified conditions. The key highlights of our investigation are succinctly outlined as follows:
\begin{enumerate} 
    \item In this investigation, we have focused on the linear form of $f(Q)$ (see Eq. \eqref{15}) model with three different forms of redshift function.
    \item It is found that the choice of redshift functions like $\phi(r)=\log\left(1+\frac{r_0}{r}\right)$ and $\frac{1}{r}$ do not satisfy the asymptotic criteria of the shape function and hence they are not suitable for this study when working under the influence of global monopole charge (see Subsubsections \ref{subsubsec2}- \ref{subsubsec3} and \ref{subsubsec5}- \ref{subsubsec6}). Thus, we discuss the below conclusion on the basis of the constant redshift function only.
    \item The shape function derived under the linear barotropic EoS case demonstrates adherence to the asymptotic flatness condition $\frac{b(r)}{r}\to 0$ as $r \to \infty$ and the flaring-out condition $b'(r_0)<1$ for the EoS parameter $\omega$ and Global Monopole parameter $\eta$, as depicted through contour plots \ref{fig1}-\ref{fig2}. 
    \item Furthermore, the shape function derived under the anisotropic EoS case illustrates adherence to the asymptotic flatness and flaring-out conditions concerning the parameters $n$ and $\eta$. This adherence is visually represented through Figures \ref{fig7}-\ref{fig8}.
    \item These observations underscore the consistency and applicability of the calculated shape functions within these theoretical frameworks, indicating the viability of our solutions within the context of wormhole physics.
    \item Embedding diagrams for the calculated shape functions given by Eqs. \eqref{21} and \eqref{27} under both EoS cases are presented in Figs. \ref{fig13} and \ref{fig14}.
    \item Analysis of the energy density, as depicted in Figs. \ref{fig3} and \ref{fig9} reveal a positively decreasing behavior throughout space-time under both linear EoS cases.
    \item Examination of Figures \ref{fig4}-\ref{fig5} and \ref{fig10}-\ref{fig11} indicates violation of the NEC for radial pressure while it is satisfied for tangential pressure. Consequently, both the NEC and WEC are violated overall, suggesting the potential presence of exotic matter at the wormhole throat to keep the throat open. Additionally, it is noteworthy that the SEC (Refer Figs. \ref{fig6} and \ref{fig12}) is fulfilled for barotropic and anisotropic EoS cases.
    \item Notably, in the isotropic EoS case, the calculated shape function fails to satisfy the asymptotic flatness condition, i.e., the ratio $\frac{b(r)}{r}$ does not tend to $0$ as $r \to \infty$. Consequently, obtaining asymptotically flat wormhole solutions for isotropic pressure under a constant redshift function for the $f(Q)$ model given in Eq. \eqref{15} becomes challenging.
    \item Table \ref{table:1} contains details regarding the energy conditions for different EoS cases with constant redshift function. It is assumed that the wormhole throat is situated at $r_0=1$.
    \item In Section \ref{sec5}, we utilized a Tolman-Oppenheimer-Volkov tool to assess the stability of the wormhole. A notable observation from graphs \ref{fig15}-\ref{fig16} is that hydrostatic and anisotropic forces exhibit similar patterns despite their opposite directions. This symmetrical balance between the forces suggests the stability of the wormhole solutions obtained in our study.
    \item  In addition, we maintain other parameters at constant values, including $\alpha=-2$, $\beta=0.02$, and $r_0 = 1$.
\end{enumerate}
In conclusion, our derived solutions for the barotropic and anisotropic EoS cases are deemed physically viable within the $f(Q)$ gravity framework with a very small contribution of global monopole charge. As a potential avenue for future research, extending this analysis to explore other modified theories of gravity could offer valuable insights and contribute to a deeper understanding of gravitational dynamics in diverse astrophysical scenarios. Moreover, exploring wormhole solutions within the framework of $f(Q)$ gravity, considering non-constant redshift functions and incorporating various matter sources, would be an intriguing avenue for future research.\\

\section*{Data Availability}
There are no new data associated with this article.

\acknowledgments  MT acknowledges the University Grants Commission (UGC), New Delhi, India, for awarding the National Fellowship for Scheduled Caste Students (UGC-Ref. No.: 201610123801). PKS acknowledges National Board for Higher Mathematics (NBHM) under the Department of Atomic Energy (DAE), Govt. of India, for financial support to carry out the Research project No.: 02011/3/2022 NBHM(R.P.)/R\&D II/2152 Dt.14.02.2022.


\begin{thebibliography}{52}
\footnotesize
\bibitem{L. Flamm} L. Flamm, \textit{Phys. Z.} \textbf{17}, 448 (1916).
\bibitem{A. Einstein} A. Einstein and N. Rosen, \textit{Phys. Rev.} \textbf{48}, 73 (1935).
\bibitem{R. W. Fuller} R. W. Fuller and J. A. Wheeler, \textit{Phys. Rev.} \textbf{128}, 919 (1962).
\bibitem{M. S. Morris 1} M. S. Morris, K. S. Thorne, and  U. Yurtsever, \textit{Phys. Rev. Lett.} \textbf{61}, 1446 (1988).
\bibitem{M. S. Morris} M. S. Morris and K. S. Thorne, \textit{Am. J. Phys.} \textbf{56}, 395 (1988).
\bibitem{M. Visser 1} M. Visser, \textit{Lorentzian Wormholes: From Einstein to Hawking} (American Institute of Physics, New York, 1996).
\bibitem{J. A. Wheeler} J. A. Wheeler, \textit{Geons. Phys. Rev.} \textbf{97}, 511 (1955).
\bibitem{P. Kanti 1} P. Kanti, B. Kleihaus, J. Kunz, \textit{Phys. Rev. Lett.} \textbf{107}, 271101 (2011).
\bibitem{P. Kanti 2} P. Kanti, B. Kleihaus, J. Kunz, \textit{Phys. Rev. D} \textbf{85}, 044007 (2012).
\bibitem{P. K. H. Kuhfittig}  P. K. H. Kuhfittig, \textit{Phys. Rev. D} \textbf{67}, 064015 (2003).
\bibitem{C. G. Bohmer 1}  C. G. B\"ohmer, T. Harko, and F. S. N. Lobo, \textit{Phys. Rev. D} \textbf{76}, 084014 (2007).
\bibitem{B. J. Barros} B. J. Barros and F. S. N. Lobo, \textit{Phys. Rev. D} \textbf{98}, 044012 (2018).
\bibitem{N. Tsukamoto} N. Tsukamoto and T. Kokubu, \textit{Phys. Rev. D} \textbf{98}, 044026 (2018).
\bibitem{Mak} T. Harko, F. S. N. Lobo, M. K. Mak, and S. V. Sushkov, \textit{ Phys. Rev. D} \textbf{87}, 067504 (2013).
\bibitem{Zangeneh} M. K. Zangeneh, F. S. N. Lobo, N. Riazi, \textit{Phys. Rev. D} \textbf{90}, 024072 (2014).
\bibitem{Galiakhmetov} K. A. Bronnikov, A. M. Galiakhmetov, \textit{Gravity Cosmol.} \textbf{21}, 283 (2015).
\bibitem{Kar2} R. Shaikh, S. Kar, \textit{Phys. Rev. D} \textbf{94}, 024011 (2016).
\bibitem{Ziaie} M. R. Mehdizadeh, A. H. Ziaie, \textit{Phys. Rev. D} \textbf{99}, 064033 (2019).
\bibitem{Singleton} V. D. Dzhunushaliev, D. Singleton, \textit{Phys. Rev. D} \textbf{59}, 064018 (1999).
\bibitem{Leon} J. P. de Leon, \textit{J. Cosmol. Astropart. Phys.} \textbf{2009}, 013 (2009).
\bibitem{Folomeev} V. Dzhunushaliev, V. Folomeev, \textit{Mod. Phys. Lett. A} \textbf{29}, 1450025 (2014).
\bibitem{V. De Falco} V. De Falco and S. Capozziello, \textit{Phys. Rev. D} \textbf{108}, 104030 (2023).
\bibitem{Karakasis} T. Karakasis, E. Papantonopoulos, and C. Vlachos, \textit{Phys. Rev. D} \textbf{105}, 024006 (2022).
\bibitem{Golchin} H. Golchin, M. R. Mehdizadeh, \textit{Eur. Phys. J. C}\textbf{ 79}, 777 (2019).
\bibitem{Eid} A. Eid, \textit{Phys. Dark Univ.} \textbf{30}, 100705 (2020).
\bibitem{Goswami} D. J. Gogoi and U. D. Goswami, \textit{J. Cosmol. Astropart. Phys.} \textbf{02}, 027 (2023).
\bibitem{Malik}  A. Malik, et al. \textit{Eur. Phys. J. C} \textbf{83}, 522 (2023).
\bibitem{Ahmad} M. Zubair, S. Waheed, Y. Ahmad, \textit{Eur. Phys. J. C} \textbf{76}, 444 (2016).
\bibitem{Bhatti} Z. Yousaf, M. Ilyas,  and M. Z. Bhatti, \textit{Eur. Phys. J. Plus} \textbf{132}, 268 (2017).
\bibitem{Chanda} A. Chanda, S. Dey, and B. C. Paul, \textit{Gen. Relativ. Grav.} \textbf{53}, 78 (2021).
\bibitem{Rosa} J. L. Rosa, P. M. Kull, \textit{Eur. Phys. J. C} \textbf{82}, 1154 (2022).
\bibitem{Tayde 1} M. Tayde, S. Ghosh, and P. K. Sahoo, \textit{Chinese Phys. C} {\bf 47}, 075102 (2023).
\bibitem{Tayde 2} M. Tayde, J. R. L. Santos, J. N. Araujo, and P. K. Sahoo, \textit{Eur. Phys. J. Plus} \textbf{138}, 539 (2023). 
\bibitem{C. G. Bohmer 2} C. G. Boehmer, T. Harko, F. S. N. Lobo, \textit{Phys. Rev. D} \textbf{85}, 044033 (2012).
\bibitem{Rani} M. Sharif and Shamaila Rani, \textit{Phys. Rev. D} \textbf{88}, 123501 (2013).
\bibitem{Momeni} M. Jamil, D. Momeni, R. Myrzakulov, \textit{Eur. Phys. J. C} \textbf{73}, 2267 (2013).
\bibitem{Tayde 3} M. Tayde, Z. Hassan and P. K. Sahoo, \textit{Phys. Dark Univ.} \textbf{42}, 101288 (2023).
\bibitem{Pavlovic} S. Bahamonde, M. Jamil, P. Pavlovic, M. Sossich, \textit{Phys. Rev. D} \textbf{94}, 044041 (2016).
\bibitem{J. B. Jimenez} J. B. Jimenez, L. Heisenberg, and T. Koivisto, \textit{Phys. Rev. D} \textbf{98}, 044048 (2018).
\bibitem{R. Lazkoz}  R. Lazkoz, et al., \textit{Phys. Rev. D} \textbf{100}, 104027 (2019).
\bibitem{S. Mandal}  S. Mandal, P. K. Sahoo, and J. R. L.Santos,
\textit{Phys. Rev. D} \textbf{102}, 024057 (2020).
\bibitem{G. Gadbail} G. Gadbail, S. Mandal, and P. K. Sahoo, \textit{Phys. Lett. B} \textbf{835}, 137509 (2022).
\bibitem{Zinnat 2} Z. Hassan, et al., \textit{Gen. Reltiv. Gravit.} \textbf{55}, 90 (2023).
\bibitem{Zinnat 3} Z. Hassan, et al., \textit{Eur. Phys. J. C} \textbf{82}, 1116 (2022).
\bibitem{L. Heisenberg} L. Heisenberg, \textit{Phys. Rept.} \textbf{1066}, 1-78 (2024).
\bibitem{F. Parsaei} F. Parsaei, et al., \textit{Eur. Phys. J. Plus} \textbf{137}, 1083 (2022).
\bibitem{O. Sokoliuk 1} O. Sokoliuk, et al., \textit{Ann. Phys.} \textbf{443}, 168968 (2022).
\bibitem{P. Bhar} P. Bhar, et al., \textit{Eur. Phys. J. C} \textbf{83}, 646 (2023)
\bibitem{S. Pradhan} S. Pradhan, et al., \textit{Chinese Phys. C} \textbf{47}, 055103 (2023).
\bibitem{O. Sokoliuk 3} O. Sokoliuk, et al., \textit{Eur. Phys. J. Plus}, \textbf{137}, 1077 (2022).

\bibitem{Debasmita} D. Mohanty, et al., \textit{Ann. Phys.} \textbf{463}, 169636 (2024).
\bibitem{Tayde 5} M. Tayde, Z. Hassan, and P. K. Sahoo, \textit{Nucl. Phys. B} \textbf{1000}, 116478 (2024).
\bibitem{P. A. M. Dirac} P. A. M. Dirac, \textit{Proc. Royal Soc. London} \textbf{A133}, 60-72  (1931).
\bibitem{G.'t Hooft} G. 't Hooft, \textit{Nucl. Phys. B} \textbf{79}, 276-284 (1974).
\bibitem{A. M. Polyakov} A. M. Polyakov, \textit{JETP Lett.} \textbf{20}, 194-195  (1974).
\bibitem{M. Barriola} M. Barriola and A. Vilenkin, \textit{Phys. Rev. Lett.} \textbf{63}, 341 (1989).
\bibitem{K. G. Zloshchastiev} K. G. Zloshchastiev, \textit{Phys. Rev. D} \textbf{57}, 4812 (1998).
\bibitem{De Chang Da} De-Chang Da and D. Stojkovic, \textit{Phys. Rev. D} \textbf{100}, 083513 (2019).
\bibitem{F. Ahmed} F. Ahmed, \textit{JCAP} \textbf{11}, 082 (2023).
\bibitem{X. Shi} X. Shi and Xin-zhou Li, \textit{Class. Quant. Grav.} \textbf{8}, 761 (1991).
\bibitem{D. P. Bennett} D. P. Bennett and S. H. Rhie,  \textit{Phys. Rev. Lett.} \textbf{65}, 1709 (1990).
\bibitem{R. H. Brandenberger} R. H. Brandenberger and H. Jiao,  \textit{JCAP} \textbf{02} 002 (2020).
\bibitem{D. Harari} D. Harari and C. Lousto,  \textit{Phys. Rev. D} \textbf{42}, 2626 (1990).
\bibitem{M. Yamaguchi} M. Yamaguchi,  \textit{Phys. Rev. D } \textbf{64},  081301 (2001).
\bibitem{A. Vilenkin} A. Vilenkin,  \textit{Phys. Rev. Lett.} \textbf{72},  3137 (1994).
\bibitem{R. Basu} R. Basu, A. H. Guth, and A. Vilenkin, \textit{Phys. Rev. D} \textbf{44}, 340 (1991).
\bibitem{R. Durrer1} R. Durrer and Z. H. Zhou,  \textit{Phys. Rev. D} \textbf{53}, 5394 (1996).
\bibitem{R. Durrer2} R. Durrer, M. Kunz, and A. Melchiorri, 
\textit{Phys. Rep.} \textbf{364}, 1-81  (2002).
\bibitem{M. Kalam} M. Kalam and P. Das,  \textit{Eur.
Phys. J. C} \textbf{82}, 342 (2022).
\bibitem{Rahaman11} F. Rahaman et al. \textit{Eur. Phys. J. C} \textbf{83}, 395 (2023).
\bibitem{S. Sarkar} S. Sarkar, N. Sarkar, and F. Rahaman, \textit{Eur. Phys. J. C} \textbf{80}, 882 (2020).
\bibitem{P. Das} P. Das and M. Kalam, \textit{Eur. Phys. J. C} \textbf{82}, 342 (2022).
\bibitem{K. Jususfi} K. Jusufi, \textit{Phys. Rev. D} \textbf{98}, 044016 (2018).
\bibitem{Tayde 4}  M. Tayde, Z. Hassan, P. K. Sahoo, and S. Gutti, \textit{Chinese Physics C} \textbf{46}, 115101 (2022).
\bibitem{R solanki} R. Solanki, A. De, and P. K. Sahoo, \textit{Phys. Dark Univ.} \textbf{36}, 101053 (2022).
\bibitem{Avik De} Avik De and Tee-How Loo, \textit{Class. Quantum Grav.}, \textbf{40}, 115007 (2023).
\bibitem{G. Mustafa} G. Mustafa, Z. Hassan, and P. K. Sahoo, \textit{Ann. Phys.} \textbf{437}, 168751 (2022).
\bibitem{S V Lohakare} S. V. Lohakare, et al., \textit{MNRAS} \textbf{526}, 3796-3814 (2023).
\bibitem{S. K. Maurya} S. K. Maurya, et al., \textit{Eur. Phys. J. C} \textbf{83}, 317 (2023).
\bibitem{A. Ditta} A. Ditta, et al., \textit{Eur. Phys. J. C} \textbf{83}, 254 (2023).
\bibitem{A.  Errehymy} A.  Errehymy, et al., \textit{Eur. Phys. J. Plus} \textbf{137}, 1311  (2022).
\bibitem{F. S. N. Lobo 1} F. S. N. Lobo, \textit{Phys. Rev. D} \textbf{71}, 084011 (2005).
\bibitem{F. S. N. Lobo 2} F. S. N. Lobo, F. Parsaei, and N. Riazi, \textit{Phys. Rev. D} \textbf{87}, 084030 (2013).
\bibitem{Z. Hassan 1} Z. Hassan, S. Mandal, and P. K. Sahoo, \textit{Forts. Phys.} \textbf{69}, 2100023 (2021).
\bibitem{K. N. Singh} K. N. Singh, A. Banerjee, F. Rahaman, and M. K. Jasim \textit{Phys. Rev. D} \textbf{101}, 084012 (2020).
\bibitem{R. Solanki} R. Solanki, Z. Hassan, and P. K. Sahoo, \textit{Chin. J. Phys} \textbf{85}, 74-88 (2023).
\bibitem{N. Godani} N. Godani and G. C. Samanta, \textit{Eur. Phys. J. C} \textbf{80}, 30 (2020).
\bibitem{S. Kar} S. Kar and D. Sahdev, \textit{Phys. Rev. D} \textbf{52}, 2030 (1995).
\bibitem{F. Rahaman 1} F. Rahaman, M. Kalam, M. Sarker, A. Ghosh and B. Raychaudhuri, \textit{, Gen. Relativ. Gravit.} \textbf{39}, 145 (2007).
\bibitem{P. H. R. S. Moraes} P. H. R. S. Moraes and P. K. Sahoo, \textit{Phys. Rev. D} \textbf{96}, 044038 (2017).
\bibitem{Oppenheimer} J. R. Oppenheimer and G. M. Volkoff, \textit{Phys. Rev.} \textbf{55}, 374 (1939).
\bibitem{Gorini} V. Gorini, U. Moschella, A. Y.  Kamenshchik, V. Pasquier and A. A. Starobinsky, \textit{Phys. Rev. D} \textbf{78}, 064064 (2008).
\bibitem{Kuhfittig} P. K. F. Kuhfittig, \textit{Fundamental J. Mod. Phys.} \textbf{14}, 23-31 (2020).
\end{thebibliography}
\end{document}